\documentclass[]{JHEP3}

\usepackage{graphicx}
\usepackage{graphics}
\usepackage{epsfig}
\epsfclipon

\usepackage{epsfig,bm}
\usepackage{latexsym}
\usepackage{amssymb,amsmath}

\newcommand{\nco}{\newcommand}
\nco{\beq}{\begin{equation}} \nco{\eeq}{\end{equation}}
\nco{\beqa}{\begin{eqnarray}} \nco{\eeqa}{\end{eqnarray}}

\def\be{\begin{equation}}
\def\ee{\end{equation}}

\def\baray{\begin{eqnarray}}
\def\earay{\end{eqnarray}}

\nco{\sss}{\scriptscriptstyle} \nco{\dphi}{\varphi}
\nco{\lsim}{\mbox{\raisebox{-.6ex}{~$\stackrel{<}{\sim}$~}}}
\nco{\gsim}{\mbox{\raisebox{-.6ex}{~$\stackrel{>}{\sim}$~}}}

\def\IK{\relax{\rm I\kern-.20em K}}
\def\IM{\relax{\rm I\kern-.20em M}}

\def\lsim{\mbox{\raisebox{-.6ex}{~$\stackrel{<}{\sim}$~}}}
\def\gsim{\mbox{\raisebox{-.6ex}{~$\stackrel{>}{\sim}$~}}}
\def\sss{\scriptscriptstyle}

\def\done{\delta^{(1)}}
\def\dtwo{\delta^{(2)}}

\def\grad{\vec{\nabla}}

\title{Dynamics with Infinitely Many Derivatives: The Initial Value Problem}

\author{Neil Barnaby \\ Canadian Institute for Theoretical Astrophysics,
University of Toronto, 60 St.\ George St.\, Toronto, Ontario M5S 3H8 Canada \\
Email: \email{barnaby@cita.utoronto.ca}}

\author{Niky Kamran \\ Department of Mathematics and Statistics, McGill University,
Montr\'eal, Qu\'ebec, H3A 2K6 Canada \\ Email: \email{nkamran@math.mcgill.ca}}

\preprint{} \abstract{ Differential equations of infinite order are
an increasingly important class of equations in theoretical physics.
Such equations are ubiquitous in string field theory and have
recently attracted considerable interest also from cosmologists.
Though these equations have been studied in the classical
mathematical literature, it appears that the physics community is
largely unaware of the relevant formalism. Of particular importance
is the fate of the initial value problem. Under what circumstances
do infinite order differential equations possess a well-defined
initial value problem and how many initial data are required?  In
this paper we study the initial value problem for infinite order
differential equations in the mathematical framework of the formal
operator calculus, with analytic initial data.  This formalism
allows us to handle simultaneously a wide array of different
nonlocal equations within a single framework and also admits a
transparent physical interpretation. We show that  differential
equations of infinite order do not generically admit infinitely many
initial data.  Rather, each pole of the propagator contributes two
initial data to the final solution.  Though it is possible to find
differential equations of infinite order which admit well-defined
initial value problem with only two initial data, neither the
dynamical equations of $p$-adic string theory nor string field
theory seem to belong to this class. However, both theories can be
rendered ghost-free by suitable definition of the action of the
formal pseudo-differential operator. This prescription restricts the
theory to frequencies within some contour in the complex plane and
hence may be thought of as a sort of ultra-violet cut-off.  Our
results place certain recent attempts to study inflation in the
context of nonlocal field theories on a much firmer mathematical
footing.  }


\begin{document}

\section{Introduction}

\subsection{Differential Equations of Infinite Order in Theoretical Physics}

Differential equations containing an infinite number of derivatives (both time and space derivatives) are an
increasingly important class of equations in theoretical physics.  Nonlocal field theories with infinitely many
powers of the d'Alembertian operator $\Box$ (given by $\Box = -\partial_t^2 + \grad^2$ in flat space)
appear ubiquitous in string field theories \cite{sft1}-\cite{pressure} (for a review see \cite{sft_review}).  This nonlocal structure
is also shared by $p$-adic string theory \cite{padic_st} (see also \cite{zwiebach}), a toy model of the bosonic string tachyon.
Yet another example of a field theory containing infinitely many powers of $\Box$ can be obtained by quantizing strings on a random
lattice \cite{random_lattice} (see also \cite{ghoshal}).  Moreover, field theories containing infinitely many derivatives have recently received considerable attention
from cosmologists \cite{phantom1}-\cite{nongaus} due to a wide array
of novel cosmological properties including the possibility of realizing quintessence with $w < -1$ within a sensible microscopic theory \cite{phantom1}-\cite{null},
improved ultra-violet (UV) behaviour \cite{tirtho1,tirtho2}, bouncing solutions \cite{tirtho1}-\cite{bounce} and self-inflation \cite{justin} (see also \cite{nl_cosmo}
for applications both to bouncing cosmologies and also to dark energy and see \cite{cosmol_tac,Joukovskaya} for a discussion of the construction of cosmological solutions in infinite order theories).
In \cite{pi} it was shown that cosmological models based on $p$-adic string theory can give rise
to slow roll inflation even when the potential is extremely steep.  This remarkable behaviour was found to be a rather general feature of nonlocal hill-top inflationary models in \cite{lidsey}.
In \cite{nongaus} it was shown that nonlocal hill-top inflation is among the very few classes of inflationary models which can give rise to a large nongaussian
signature in the Cosmic Microwave Background (CMB).

Differential equations with infinitely many derivatives which arise frequently in the literature include the dynamical equation of $p$-adic string theory \cite{padic_st}
\begin{equation}
\label{padic}
  p^{-\Box / 2} \phi = \phi^p
\end{equation}
where $p$ is a prime number (though it appears that the theory can be sensibly continued to other values of $p$ also \cite{p=1}) and we have set $m_s \equiv 1$.
A second popular example is the dynamical equation of the tachyon field in bosonic open string field theory (SFT) which can be cast in the form (see, for example, \cite{sen:padicdescent})
\begin{equation}
\label{SFT_intro}
  \left[ \left(1 + \Box \right) e^{-c \Box}  - 2\right] \phi = \phi^2
\end{equation}
where $c = \ln (3^3 / 4^2 )$.  In both cases the field $\phi$ is a
tachyon representing the instability of some non BPS D-brane
configuration.  More generally, there is considerable interest in applications of a wide class of equations of the form \cite{bounce,gen_nonlocal,route,zeta}
\begin{equation}
\label{gen_intro}
  F\left(\Box \right) \phi = U\left( \phi \right)
\end{equation}
where $U(\phi) = V'(\phi)$ is the derivative of some potential energy function $V(\phi)$ associated with the field $\phi$ and we refer to $F(z)$ as the \emph{kinetic function}.
Equations of the form (\ref{gen_intro}) are interesting in their own right from the mathematical perspective and some special cases have received attention recently
\cite{padic_math}-\cite{vlad}.

Somewhat more general classes of infinite order differential equations, in which the derivatives do not necessarily appear in the combination $\Box = -\partial_t^2 + \grad^2$,
arise in noncommutative field theory \cite{ncft}, fluid dynamics \cite{kdv,fluid} and quantum algebras \cite{kdv}.

\subsection{The Importance of the Initial Value Problem}

Of particular interest is the fate of the initial value problem for
infinite derivative theories.  When does equation (\ref{gen_intro})
admit a well-defined initial value problem - even formally, that is
ignoring issues of convergence - and how many initial data are
required to determine a solution? Such questions are fundamental to
any physical application.  To emphasize the nature of the problem that we are solving,
let us consider a few trivial examples.  First consider the equation
\begin{equation}
\label{example1}
  (\partial_t^2 - m^2)^2 \, \phi = 0
\end{equation}
It is easy to see that, due to the presence of the operator $\partial_t^2 - m^2$, the function
\begin{equation}
\label{example1_soln1}
  \phi_1(t) = A e^{mt} + B e^{-mt}
\end{equation}
provides a two-parameter solution of (\ref{example1}).  Of course, this is not the most general solution.  To $\phi_1$ one could also append
\begin{equation}
\label{example1_soln2}
  \phi_2(t) = t \left [C e^{mt} + D e^{-mt} \right]
\end{equation}
Since equation (\ref{example1}) is fourth order in derivatives it is not surprising that the four parameter solution $\phi(t) = \phi_1(t) + \phi_2(t)$ provides
the most general possible solution and the initial value problem is well-posed with four initial conditions.  We now consider the somewhat less trivial example
\begin{equation}
\label{example}
  \mathrm{Ai}(-\partial_t^2) \, \phi(t) = 0
\end{equation}
where $\mathrm{Ai}(z)$ is the Airy function.  It is perhaps obvious that a solution is provided by
\begin{equation}
\label{ex_soln}
  \phi(t) = \sum_{n} \left[ a_n e^{m_n t} + b_n e^{-m_n t} \right]
\end{equation}
where $\{ m_n^2 \}$ are the zeroes of the $\mathrm{Ai}(-z)$ (the
first few roots are $m_1^2 \cong 1.17$, $m_2^2 \cong 3.27$, $\cdots$
). What is probably less obvious is whether (\ref{ex_soln}) provides
\emph{the most general solution} of (\ref{example}).  Could there be
additional solutions which are not as trivial to construct as
(\ref{ex_soln})?  Since the original equation (\ref{example}) was
infinite order, it is not entirely obvious.  An even less trivial
example is provided by the equation
\begin{equation}
\label{example2}
  \sqrt{\partial_t^2 - m^2}\, \phi = 0
\end{equation}
which has appliciations to bosonization.  How many
initial conditions are necessary to specify a solution of (\ref{example2})?

In this paper we study the initial value problem for differential
equations with infinitely many derivatives in the context of the
formal operator calculus.  The initial value problem for various
nonlocal theories has also been studied in \cite{route},
\cite{ostrogradski}-\cite{maximal}. In this paper we lay down a
simple and intuitive formalism for studying the initial value
problem in nonlocal theories which is sufficiently general to handle
a wide variety of different types of nonlocality.
(In particular, our formalism is sufficiently general to handle the motivational
examples (\ref{example1}, \ref{example}, \ref{example2}).)
Though we do not
derive any new solution of (\ref{padic}, \ref{SFT_intro}) we
nevertheless feel that it is instructive to re-consider such
equations in the context of our formalism.  Our approach may be
readily applied also to other nonlocal equations which arise in
physical applications.

It is not uncommon to see equations of the form (\ref{gen_intro})
described as ``a new class of equations in mathematical physics'' in
the string theory and cosmology literature. However, it happens that
differential equations of infinite order have been studied in the
mathematical literature for quite some time. Indeed, the study of
linear differential equations of infinite order was the subject of
an extensive treatise by H.\ T.\  Davis as early as 1936
\cite{davis}! This treatise and papers by Davis \cite{davis1} and
Carmichael \cite{carmi} give an account of this theory as it stood
at the time. The topic was further developed from an analytical
perspective by Carleson \cite{Carleson}, who showed that the
solutions of differential equations of infinite order need not be
analytic functions, and obtained sufficient conditions in terms of
the coefficients of the equation for the solutions of the initial
value problem to be analytic. We should also mention that initial
value problems for some special classes of differential equations of
infinite order which are not of the type studied in this paper
appear in the modern theory of pseudo-differential operators
\cite{Taylor}. However, for the purposes of this paper, the symbolic
calculus described in the classical papers \cite{davis} and
\cite{carmi} will be sufficient, since our main focus is on the
determination of the number of parameters on which the analytic
local solutions of the initial value problem could possibly depend.
It is a bit surprising that the physics community has
apparently not been aware of this classical mathematical literature,
given the simplicity of the mathematical formalism which relies only
on some basic results from complex analysis and the theory of
integral transforms.  One of our primary goals in the current note is to
bring this mathematical literature to the attention of the physics
community and to apply the formalism to certain equations of
particular physical interest.

Since differential equations of $N$-th order (in the time
derivative) require $N$ initial data it is sometimes reasoned that
equations of the form (\ref{gen_intro}) admit a unique solution only
once infinitely many initial data are specified (as in our toy
example (\ref{example})).  Such a situation would pose serious
difficulties for any physical application: by suitable choice of
infinite data one could freely specify the solution $\phi(t)$ to
arbitary accuracy in any finite time interval $\Delta t$
\cite{woodard2}.  In such a situation the initial value problem
completely looses predictivity. Fortunately, it is not generically
the case that differential equations of infinite order admit
infinitely many initial data.
We will show that for free field theories every pole of the propagator contributes two
initial data to the solution of the field equation. This result is
simple to understand on physical grounds since each pole of the
propagator corresponds to a physical excitation in the theory and,
on quantization, the two initial data per degree of freedom are
promoted to annihilation/creation operators.

In the context of quantum field theory the question of counting
initial data is intimately related to the question of whether the
theory suffers from the presence of ghosts - quantum states having
wrong-sign kinetic term in the Lagrangian. The presence of ghosts
signals a pathology in the underlying quantum field theory since
these states either violate unitarity or else carry negative
kinetic energy and lead to vacuum instability \cite{jim}. One must worry
about the presence of ghosts in theories which give rise to
equations of the form (\ref{gen_intro}) since the addition of
\emph{finitely} many higher derivative terms in the Lagrangian
generically introduces ghost-like excitations into the theory.  As
we shall see later on, this is not necessarily the case in
nonlocal field theories containing \emph{infinitely} many higher
derivative terms.  A related worry is the presence of the
Ostrogradski instability \cite{ostrogradski} (see
\cite{woodard1,woodard2} for a review) which generically plagues
\emph{finite} higher derivative theories.  The Ostrogradski
instability is essentially the classical manifestation of having
ghosts in the theory.  Later on we will elucidate more carefully
the relationship between these issues.

Though it is possible to constuct nonlocal equations which evade the Ostrogradski
instability, we will show that neither the dynamical equations of $p$-adic string theory (\ref{padic})
nor SFT (\ref{SFT_intro}) belong to this class.  However, we illustrate a simple nonperturbative prescription
for re-defining these theories in such a way as to evade such difficulties.  This prescription restricts the theory
to frequencies $\omega$ within some contour $C$ in the complex plane and may be analogous to placing an ultra-violet (UV) cut-off on the theory.

The plan of this paper is as follows.  In section \ref{ode_sec} we
consider linear ordinary differential equations of infinite order, laying
down the necessary formalism for counting initial conditions. In
section \ref{pde_sec} we generalize this analysis to partial
differential equations of infinite order (in the case  that the
derivatives appear in the combination $\Box = -\partial_t^2 +
\grad^2$) showing that the initial data counting has a transparent
physical interpretation and commenting also on the failure of the
Ostrogradski construction.  In section \ref{examples_sec} we apply
our prescription for initial data counting to some particular
infinite order differential equations which appear frequently in
the literature.  In section \ref{nonlin_sect} we discuss the nonlinear problem and
illustrate the application of our formalism to nonlinear equations by studying
the equation of $p$-adic string theory using a perturbative
expansion about the unstable vacuum of the theory.  We present our
conclusions in section \ref{concl}.  Appendix A reviews the relation between
(\ref{padic}) and a certain nonlinear Fredholm equation; appendix
B gives some technical details concerning our conventions for the
Laplace transform and appendix C applies our initial data counting
to open-closed $p$-adic string theory.

\section{Linear Ordinary Differential Equations of Infinite Order: Counting the Initial Data}
\label{ode_sec}

Before specializing to particular equations of immediate physical interest
we first develop the general theory of linear ordinary differential
equations (ODEs) of the form
\begin{equation}
\label{general_eqn}
  f(\partial_t)\phi(t) = J(t)
\end{equation}
where the function $f(s)$ is often called the \emph{generatrix} in
mathematical literature. Equations of the form (\ref{general_eqn})
are closely associated with both Fredholm integral equations and this relation is illustrated in appendix A for the
case of the $p$-adic string.  There
are various prescriptions which one might use to define the action
of the formal pseudo-differential operator $f(\partial_t)$. In the
case that $f(s)$ is an analytic function one might represent it by
the convergent series expansion
\begin{equation}
\label{generatrix_series}
  f(s) = \sum_{n=0}^{\infty}\frac{f^{(n)}(0)}{n!} \, s^{n}
\end{equation}
so that (\ref{general_eqn}) can be written as
\[
  \sum_{n=0}^{\infty} \frac{f^{(n)}(0)}{n!}\partial_t^{(n)} f(t) = J(t)
\]

In the case where $f(s)$ is non-analytic
however, some alternative definition is required.
(Notice that the possibility of non-analytic generatrix is not purely academic.  The zeta-strings model of \cite{zeta} involves a nonanalytic kinetic function.)
A natural prescription is to define the pseudo-differential operator $f(\partial_t)$
through its action on Laplace transforms.\footnote{One might instead imagine using a Laurent series definition, however,
such a definition would be inherently ambiguous because $f(s)$ would admit different Laurent series expansions, each valid in
a different annulus about $s=0$.}  Let us assume that we are given a function
$\phi(t)$ which admits a Laplace transform $\tilde{\phi}(s)$ defined by
\begin{equation}
\label{lap_x_form}
  \phi(t) = \frac{1}{2\pi i}\oint_C ds \, e^{st} \tilde{\phi}(s)
\end{equation}
valid for $t \geq 0$\, where $C$ is a contour to be specified (see
appendix B for details\footnote{One could require for example that
$\phi(t)$ be entire, of exponential type.}). For $t < 0$ this
expression does not, strictly, apply and $\phi(t)$ should be taken
to vanish (see again appendix B for more details).  It is natural to define
$f(\partial_t) \phi(t)$ by
\begin{equation}
\label{pseudo}
  f(\partial_t)\phi(t) = \frac{1}{2\pi i}\oint_C ds \, e^{st} \left[ f(s) \tilde{\phi}(s) \right]
\end{equation}
where we drop the additional term involving $\phi^{(i)}(0)$ (see appendix B) which can be absorbed
into the arbitrary coefficients in the solutions of (\ref{general_eqn}) (this will become clear shortly).
If we make the natural choice of taking $C$ to be an contour which encloses all the poles of the integrand
in (\ref{lap_x_form}) then (\ref{pseudo}) reproduces the infinite series definition (\ref{generatrix_series})
for $f(s)$ analytic.  Moreover, this definition reproduces the one used in \cite{analysis}.
This is the choice of $C$ which is mathematically well-motivated and we will make this
choice for the most part.  However, later on we will consider a particular alternative choice of $C$ which is motivated by physical considerations.

We will be particularly interested in equations of the form
(\ref{general_eqn}) for which the generatrix may be cast in the
form
\begin{equation}
\label{generatrix_ansatz}
  f(s) = \gamma(s) \prod_{i=1}^{M} (s-s_i)^{r_i}
\end{equation}
with $\gamma(s)$ being everywhere non-zero, which implies that
$f(s)$ has precisely $M$ zeros at the points $s=s_i$, the $i$-th
zero being of order $r_i$. For simplicity we assume that all the
$r_i$ are positive integers, though this assumption is relaxed
in subsection \ref{frac_ops_subsec}. We further assume that $|s_1| < |s_2| <
\cdots < |s_M|$.  The function $\gamma(s)^{-1}$ is nonsingular on
the disk $|s| < |s_M|$ but is otherwise arbitrary.
It is useful also to introduce the \emph{resolvent generatrix} $f(s)^{-1}$
which has simple poles at the points $s=s_i$, the $i$-th pole
being of order $r_i$.

\subsection{The Homogeneous Equation}
\label{hom_subsec}

We first assume that we are given a solution of (\ref{general_eqn}) in the case that $J(t) = 0$,
and that this solution admits a Laplace transform $\tilde{\phi}(s)$ defined by (\ref{lap_x_form}).
Substituting (\ref{lap_x_form}) into (\ref{general_eqn}), yields
\begin{equation}
\label{lap_space_eqn}
  \frac{1}{2\pi i} \oint_C ds \, e^{st} f(s)\tilde{\phi}(s) = 0
\end{equation}
What is the most general function $\tilde{\phi}(s)$ which satisfies
this condition? From the Cauchy integral theorem we know that this
condition will be satisfied only if the integrand is analytic
everywhere in a neighborhood of the interior of $C$ so that
$\tilde{\phi}(s)$ may have simple poles at the points $s=s_i$, the
$i$-th pole being of order $r_i$ or less. Thus the most general
solution can be written in the form
\begin{equation}
\label{hom_soln_lap_space}
  \tilde{\phi}(s) = \frac{1}{\gamma(s)} \sum_{i=1}^{M}\sum_{j=1}^{r_i}\frac{A^{(i)}_{j}}{(s-s_i)^{j}}
\end{equation}
The factor of $\gamma(s)$ is included for convenience and an additive analytic function (which would not alter the configuration-space solution $\phi(t)$)
has been omitted.\footnote{The additional terms excluded from (\ref{pseudo}) (see appendix B) can be absorbed into the definition of $A_j^{(i)}$.}
The solution (\ref{hom_soln_lap_space}) contains a total of $N$ arbitrary coefficients $A^{(i)}_{j}$ where
\begin{equation}
\label{N}
  N = \sum_{i=1}^{M} r_i
\end{equation}
In other words, $N$ counts the zeros of $f(s)$ according to their
multiplicity.  The $N$ free coefficients $A^{(i)}_j$ will
ultimately serve to fix $N$ initial conditions for the initial
value problem corresponding to (\ref{general_eqn}).  We now insert
the solution (\ref{hom_soln_lap_space}) into (\ref{lap_x_form})
and perform the $ds$ integration (with $C$ a contour which
encloses all of the points $\{s_i\}$) using the Cauchy integral
formula
\[
  \frac{1}{2\pi i}\oint_C ds \, \frac{h(s)}{(s-s_i)^j} = \frac{1}{(j-1)!} h^{(j-1)}(s_i)
\]
(valid for $h(s)$ analytic inside $C$ and $j > 0$).  The resulting solution $\phi(t)$ in configuration space takes the form
\begin{equation}
\label{hom_soln}
  \phi(t) = \sum_{i=1}^{M} P_i(t) e^{s_i t}
\end{equation}
where each $P_i(t)$ is a polynomial of order $r_i-1$
\begin{equation}
\label{P_i}
  P_i(t) = \sum_{j=1}^{r_i} p_{j}^{(i)} t^{j-1}
\end{equation}
The $N$ coefficients $\{p_j^{(i)}\}$ are arbitrary (reflecting the fact that $A^{(i)}_{j}$ were arbitrary) and will serve to fix $N$ initial conditions $\phi^{(n)}(0)$ for
$n = 0, \cdots, N-1$.

\subsection{The Particular Solution of the Inhomogeneous Equation}
\label{par_subsec}

Having determined the solution of the homogeneous equation
(\ref{general_eqn}) we now consider the situation where $J(t) \not=
0$ and focus on the particular solution due to the source $J(t)$. We
assume that the source term has a Laplace transform $\tilde{J}(s)$
defined by
\begin{equation}
\label{J_tilde}
  J(t) = \frac{1}{2\pi i}\oint_C ds \, e^{st} \tilde{J}(s)
\end{equation}
and we continue to employ the Laplace transform $\tilde{\phi}(s)$
(\ref{lap_x_form}). Then equation (\ref{general_eqn}) takes the form
\begin{equation}
\label{inhom_lap_space_eqn}
  \frac{1}{2\pi i}\oint_C ds \, \left[ f(s)\tilde{\phi}(s) - \tilde{J}(s) \right] = 0
\end{equation}
so that the particular solution is $\tilde{\phi}(s) = \tilde{J}(s) / f(s)$ or, in configuration space
\begin{equation}
\label{inhom_soln}
  \phi(t) = \frac{1}{2\pi i}\oint_C ds\, e^{st} \frac{\tilde{J}(s)}{f(s)}
\end{equation}
From (\ref{inhom_soln}) it is clear that the resolvent generatrix is the Laplace-space Green function $\tilde{G}(s) = f(s)^{-1}$.

\subsection{The Initial Value Problem}
\label{consequence_subsec}

It is natural to expect that in order to obtain the most general
solution of (\ref{general_eqn}) we should append to
(\ref{inhom_soln}) the solution of the homogeneous equation
(\ref{hom_soln}).  One may prove the following theorem, which we now
state without proof but whose verisimilitude should be clear from
the preceding analysis (for a detailed and elegant proof, see
\cite{carmi}).

{\bf Theorem:}  { \it Consider the linear differential equation of infinite order (\ref{general_eqn}) with the function $f(s)$ analytic within
the region $|s| \leq q$ where $q$ is a given positive constant or zero and further suppose that $f(s)$ may be written in the form (\ref{generatrix_ansatz}).
If $J(t)$ is a function of exponential type not exceeding $q$ then the most general solution $\phi(t)$, subject to the condition that it shall be a
function of exponential type not exceeding $q$, may be written in the form }
\begin{equation}
\label{soln}
  \phi(t) =  \frac{1}{2\pi i}\oint_C ds\, e^{st} \frac{\tilde{J}(s)}{f(s)} + \sum_{i=1}^{M} P_i(t) e^{s_i t}
\end{equation}
{\it where $P_i(t)$ is an arbitrary polynomial of order $r_i-1$ and $\tilde{J}(s)$ is the Laplace-space source.
The first term in (\ref{soln}) corresponds to the particular solution due to the source $J(t)$ and the second term is the solution of
the homogeneous equation.  If the generatrix $f(s)$ does not vanish inside $|s| \leq q$ then the latter solution is identically zero. }

Armed with the solution (\ref{soln}) the interpretation of the initial value problem for (\ref{general_eqn}) is clear: the solution contains $N$ arbitrary coefficients
$p_j^{(i)}$ which serve to fix $N$ initial conditions $\phi^{(n)}(0)$ for $n=0,\cdots ,N-1$.  This result has a transparent physical interpretation in terms of poles of the
propagator, which we shall return to in the next section.  It is important to note that the significance of this theorem is that (\ref{soln}) provides \emph{the most general}
solution of (\ref{general_eqn}).  Returning to the motivational example (\ref{example}) we see that (\ref{ex_soln}) is, indeed, the full solution because the zeros of the Airy function
are order unity.

An interesting consequence of this theorem, whose significance we shall return to later on, is that for $J(t) = 0$ the solution (\ref{soln}) is identical to the
solution of the equation $\bar{f}(\partial_t) \phi(t) = 0$ where
\[
  \bar{f}(s) = \frac{f(s)}{\gamma(s)} = \prod_{i=1}^{M} (s-s_i)^{r_i}
\]
In other words, for the homogeneous equation the dynamics are completely insensitive to the choice of $\gamma(s)$ and the solutions are completely
determined by the pole structure of the resolvent generatrix.  For finite $M$ this implies that the dynamics of the full infinite order differential equation (provided it is linear
and source-free) will be identical to some finite order differential equation.

It is worth commenting on the generality of our
results.  From the perspective of mathematical rigour the only
serious caveat of our analysis is the assumption that the sources
$J(t)$ and solutions $\phi(t)$ are analytic, with the asymptotic
behavior needed for their Laplace transform to exist (which is not,
in general, guaranteed). Nonanalytic solutions have been discussed
for $p$-adic string theory in \cite{vlad} though it is not entirely clear how to
interpret these solutions physically.  Naively, one might expect some components of the stress tensor
associated with such solutions to blow up at some finite time.  However, \cite{singular} showed
that a particular nonanalytic solution in SFT could sensibly be regularized to yield a well-behaved
stress tensor.

\subsection{Equations Involving Fractional Operators}
\label{frac_ops_subsec}

In writing (\ref{generatrix_ansatz}) we have assumed that the zeroes of the generatrix are of integer order (we have assumed that all of the $r_i$ are integer) and we have
excluded the case of fractional differentiation ($f(s) = s^{\alpha}$ with $\alpha$ non-integer).  In this subsection we relax those assumptions.  Though such equations are less common in
applications to string theory and cosmology,
they constitute a large class of nonlocal equations which admit well-posed initial value problem with finitely many initial conditions and may be of some interest for this reason.
The reader more interested in the equations (\ref{padic}, \ref{SFT_intro}) may wish to skip to the next section.

\subsubsection{Roots of Non-Integer Order}
\label{non_int_root}

To illustrate the effect of taking non-integer $r_i$ in (\ref{generatrix_ansatz}) let us consider the simple equation
\begin{equation}
\label{root_eqn}
  \sqrt{\partial_t^2 + m^2}\,\phi(t) = 0
\end{equation}
If we seek solutions $\phi(t)$ which admit a Laplace transform
$\tilde{\phi}(s)$, we obtain
\begin{equation}
  \frac{1}{2\pi i} \oint_C ds\, \left[ \sqrt{s^2 + m^2} \,\tilde{\phi}(s) e^{st} \right] = 0
\end{equation}
which implies that
\begin{equation}
  \tilde{\phi}(s) = \frac{A}{\sqrt{s^2 + m^2}}
\end{equation}
up to an additive analytic function which does not alter the configuration-space solution.  Taking the inverse Laplace transform using the formula
\begin{equation}
\label{lap_iden}
  \mathcal{L}^{-1}\left[ \frac{1}{\sqrt{s^2 + m^2}} \right] = J_0(mt)
\end{equation}
(where $J_\nu$ is the Bessel function of the first kind) we have
\begin{equation}
\label{1/2_soln}
  \phi(t) = A J_0(m t)
\end{equation}
So that equation (\ref{root_eqn}) admits only one initial condition.

It is also straightforward to consider
\begin{equation}
\label{root_eqn2}
  \left( \partial_t^2 + m^2 \right)^{3/2} \phi(t) = 0
\end{equation}
which implies that
\begin{equation}
  \frac{1}{2\pi i} \oint_C ds\, \left[ \left( s^2 + m^2 \right)^{3/2} \tilde{\phi}(s) e^{st} \right] = 0
\end{equation}
The general solution is
\begin{equation}
  \tilde{\phi}(s) = \frac{a }{\left( s^2 + m^2 \right)^{1/2}} + \frac{b}{\left(s^2 + m^2\right)^{3/2}}
\end{equation}
with $a,b$ arbitrary.  We must now perform the inverse Laplace transform.  The first term is trivial using (\ref{lap_iden}).  By differentiating
(\ref{lap_iden}) with respect to $m$ we find that
\begin{equation}
\label{lap_iden2}
  \mathcal{L}^{-1}\left[ \frac{1}{\left(s^2 + m^2\right)^{3/2}} \right] = \frac{t}{m} J_1(mt)
\end{equation}
so that the configuation-space solution of (\ref{root_eqn2}) is
\begin{equation}
\label{3/2_soln}
  \phi(t) = A J_0(mt) + B (mt) J_1(mt)
\end{equation}
which admits two initial conditions.

It is straightforward to handle higher order equations of the form
\[
  \left(\partial_t^2 + m^2 \right)^{n/2} \phi(t) = 0
\]
with $n$ an odd integer.  The Laplace-space solution is
\[
  \tilde{\phi}(s) = \sum_{i=0}^{(n-1)/2} \frac{A_i}{(s^2 + m^2)^{(2i+1)/2}}
\]
which contains a total of $(n+1)/2$ arbitrary coefficients.  The inverse Laplace transforms may be performed by successively differentiating (\ref{lap_iden}) with respect to $m$
and using the well-known Bessel function identity
\[
  J_\nu'(x) = -J_{\nu + 1}(x) + \frac{\nu}{x}J_\nu(x)
\]

In the tachyonic case $m^2 = -\mu^2 < 0$ one simply replaces the Bessel functions $J_\nu$ with modified Bessel functions $I_\nu$ to obtain real-valued solutions.
We find that the equation
\begin{equation}
\label{root_eqn3}
  \sqrt{\partial_t^2 - \mu^2}\phi(t) = 0
\end{equation}
has solution
\begin{equation}
\label{1/2_soln2}
  \phi(t) = A I_0(\mu t)
\end{equation}
while
\begin{equation}
\label{root_eqn4}
  \left(\partial_t^2 - \mu^2\right)^{3/2}\phi(t) = 0
\end{equation}
has solution
\begin{equation}
\label{3/2_soln2}
  \phi(t) = A I_0(\mu t) + B(\mu t) I_1(\mu t)
\end{equation}

\subsubsection{Fractional Differentiation}

Another interesting case occurs when the generatrix involves noninteger powers of the derivatives, such as $f(s) = s^{\alpha}$ with $\alpha$ some arbitrary complex number.
In this case clearly the definition (\ref{generatrix_series}) does not apply.
The attempt to assign sensible meaning to the symbol $\partial_t^{\alpha}$ for noninteger $\alpha$ has a long history in mathematical analysis and there are a
number of definitions in the literature \cite{Taylor,fractional}.  For our purposes, the most useful definition seems to be the Liouville fractional derivative.
To motivate this definition we first consider the problem of defining fractional integration.  We define ordinary integration by
\begin{eqnarray*}
  (I \phi)(t) &=& \int_{0}^t dt' \phi(t') \\
  (I^2 \phi)(t) &=& \int_{0}^t dt' (I \phi)(t') \\
  &\cdots&
\end{eqnarray*}
(The choice of lower limit of integration is purely conventional.
Different choices will lead to different definitions of fractional
differentiation.) The Cauchy formula for repeated integration gives
\begin{equation}
\label{repeat}
  (I^n \phi)(t) = \frac{1}{(n-1)!} \int_0^t dt' (t-t')^{n-1} \phi(t') dt'
\end{equation}
A natural continuation to noninteger $n$ is
\begin{equation}
\label{repeat_non}
  (I^\alpha \phi)(t) = \frac{1}{(\alpha-1)!} \int_0^t dt' (t-t')^{\alpha-1} \phi(t') dt'
\end{equation}
valid for $t>0$, $\mathrm{Re}(\alpha)>0$ and where the symbol $(\alpha-1)!$ should now be interpreted as the factorial function.\footnote{That is to say $(\alpha-1)! \equiv \Gamma(\alpha)$
(where $\Gamma$ is the gamma-function) for non-integer $\alpha$.  We do not use the gamma-function notation explicitly to avoid confusion with another function which we denote by $\Gamma(z)$
which will be introduced in the next section.}
This definition is commutative and additive $I^{\alpha}(I^{\beta}\phi) = I^{\beta}(I^{\alpha}\phi) = I^{\alpha+\beta}\phi$.  It is now natural to define fractional differentiation
by differentiating (\ref{repeat_non})
\begin{eqnarray}
  (D^{\alpha}\phi)(t) &\equiv&  \left(\frac{d}{dt}\right)^n (I^{n-\alpha}\phi)(t) \nonumber \\
  &=& \frac{1}{(n-\alpha-1)!}\left(\frac{d}{dt}\right)^n \int_0^t dt' \frac{\phi(t')}{(t-t')^{\alpha-n+1}} \label{frac_der}
\end{eqnarray}
Of course for integer $\alpha$ equation (\ref{frac_der}) reproduces the usual derivative.  This definition also has the following nice properties \cite{fractional}
\begin{eqnarray*}
  (D^{\alpha} I^{\alpha} \phi)(t) &=& \phi(t) \hspace{5mm}\mathrm{if}\hspace{5mm}\alpha>0\\
  (D^{\beta} I^{\alpha} \phi)(t) &=& (I^{\alpha-\beta} \phi)(t) \hspace{5mm}\mathrm{if}\hspace{5mm}\alpha>\beta>0\\
  D^{\alpha} t^{\beta-1} &=& \frac{(\beta-1)!}{(\beta-\alpha - 1)!} t^{\beta-\alpha - 1} \hspace{5mm}\mathrm{if}\hspace{5mm}\mathrm{Re}(\alpha)>0, \hspace{2mm}\mathrm{Re}(\beta)>0
\end{eqnarray*}

The definition (\ref{repeat_non}) acts in Laplace space as
\begin{equation}
\label{lap_frac}
  \mathcal{L} \left[D^{\alpha}\phi\right] = s^{\alpha} \tilde{\phi}(s) - \sum_{i=1}^{l} d_i s^{i-1}
\end{equation}
where $l$ is a natural number such that $l-1 < \alpha \leq l$ and the constants $\{d_i\}$ are
\begin{equation}
\label{di}
  d_i \equiv (D^{\alpha-i}\phi)(0)
\end{equation}
The result (\ref{lap_frac}) is exactly of the form (\ref{pseudo}).

\subsubsection{Fractional Differential Equations}

There is a large literature on solving equations involving the
operator $D^{\alpha}$ (\ref{frac_der}) which is on rather firm
mathematical footing (both linear and nonlinear equations have been
studied).  For illustration we consider only the simple prototype
equation
\begin{equation}
\label{frac_ode}
  (D^{\alpha} \phi)(t) - m^{\alpha} \phi(t) = J(t)
\end{equation}
We first consider the homogeneous equation $J = 0$.  Taking the Laplace transform using (\ref{lap_frac}) and solving for $\tilde{\phi}(s)$ we have
\[
  \tilde{\phi}(s) = \sum_{j=1}^{l} d_j \frac{s^{j-1}}{s^{\alpha} - m^{\alpha}}
\]
Now, inverting the Laplace transform we find the solution
\[
  \phi(t) = \sum_{j=1}^{l} d_j \phi_j(t)
\]
where
\begin{equation}
\label{fund}
 \phi_j(t) = \mathcal{L}^{-1}\left[\frac{s^{j-1}}{s^{\alpha} - m^{\alpha}}\right] = t^{\alpha-j} E^{\alpha,\alpha+1-j}\left[( m t )^{\alpha}\right]
\end{equation}
where $E_{\alpha,\beta}$ is the generalized Mittag-Leffler function (see, for example, \cite{fractional}).  The fundamental solutions $\{\phi_j\}$ satisfy
\begin{eqnarray*}
  (D^{\alpha-i}\phi_j)(0) &=& 0 \hspace{5mm}\mathrm{for}\hspace{5mm}i,j = 1, \cdots,l; i > j \\
  (D^{\alpha-i}\phi_i)(0) &=& 1 \hspace{5mm}\mathrm{for}\hspace{5mm}i = 1, \cdots, l
\end{eqnarray*}
It can be verified that in fact any summation
\begin{equation}
\label{frac_hom}
    \phi(t) = \sum_{j=1}^{l} a_j \phi_j(t)
\end{equation}
with $l$ arbitrary coefficients $\{a_j\}$ yields a solution to (\ref{frac_ode}).  We see, then, that equation (\ref{frac_ode}) admits a well-posed initial value problem with $l$ initial conditions.
In particular, for $1 < \alpha \leq 2$ equation (\ref{frac_ode}) admits two initial conditions.

It is also straightforward to consider the inhomogeneous equation, following the approach of (\ref{inhom_soln}).  In fact, one may define a fractional Green function
\begin{equation}
\label{frac_green}
  G_\alpha(t) = \frac{1}{2\pi i} \oint_C ds\, \frac{e^{st}}{s^{\alpha} - m^{\alpha}}
\end{equation}
so that the particular solution is
\begin{equation}
\label{frac_inhomo}
  \phi(t) = \int_0^t dt' G_{\alpha}(t-t') J(t')
\end{equation}
which can be proved to provide a unique solution to (\ref{frac_ode}) with $\phi(0) = 0$ \cite{fractional}.

\subsubsection{Properties of Fractional Operators}

It is worth commenting on the fact that our definition of the
operator $(\partial_t^2 + m^2)^{1/2}$ and also the
Liouville\footnote{Our comments actually apply for nearly all
definitions of fractional differentiation which are considered in
the mathematics literature.  See \cite{fractional} for further
details.} definition of the fractional derivative $D^{\alpha}$
lead to some a-priori unexpected properties, the most notable of
which is that the degrees of the fractional derivative operators
which we have defined are not, in general additive. In other
words, $D^{\alpha}$ does \emph{not} satisfy $D^{\alpha}D^{\beta} =
D^{\alpha + \beta}$ for general (non-integer) $\alpha$, $\beta$
\cite{fractional}. As simple illustration of this fact is given by
considering the function $\phi(t) = t^{-1/2}$, which satisfies
$D^{1/2}\phi = 0$, while $D^{1}\phi =\partial_t \phi =
-1/2\,t^{-3/2} \not= 0$. A similar property can be inferred in the
case of the operator $(\partial_t^2 + m^2)^{1/2}$ by noting that
(\ref{root_eqn}) does not have solutions $\cos(mt)$, $\sin(mt)$.
Indeed, it is straightforward to see that the solution
(\ref{1/2_soln}) of (\ref{root_eqn}) is \emph{not} a solution of
the equation $(\partial_t^2 + m^2) \phi(t) = 0$.

Given that, using our conventions, acting twice with the operator
$(\partial_t^2 + m^2)^{1/2}$ does not necessarily return the same
result as acting once with the operator $(\partial_t^2 + m^2)$, the
reader may wonder whether the definition that we have employed is
the most sensible one.  We would like to argue that, although the
definition we have adopted has the unpleasant property that the
degrees are in general not additive, it is the most physically
reasonable approach. To see this, we consider an alternative
definition of $(\partial_t^2 + m^2)^{1/2}$ and contrast this with
ours.  As an alternative to our approach let us consider defining
$f(\partial_t) = \sqrt{\partial_t^2 + m^2}$ by an infinite series of
the form
\begin{equation}
\sqrt{\partial_t^2 + m^2} =
\partial_{t}+Q_{0}+Q_{-1}\,\partial_{t}^{-1}+Q_{-2}\,\partial_{t}^{-2}+Q_{-2}\,\partial_{t}^{-2}+
\cdots
\end{equation}
Substituting this into the defining identity
\begin{equation}
\label{def_id}
\sqrt{\partial_t^2 + m^2}.\sqrt{\partial_t^2 + m^2} =
\partial_{t}^{2}+m^{2},
\end{equation}
and applying the formal identity
\begin{equation}
\partial_{t}^{-1}.Q =
\sum_{i=0}^{\infty}(-1)^{i}Q^{(i)}\partial_{t}^{-i-1},
\end{equation}
to solve for the coefficients $Q_{0},\,Q_{-1},Q_{-2},Q_{-3}\ldots $ we arrive at
\begin{equation}
\label{series1}
  \sqrt{\partial_t^2 + m^2} \equiv \partial_t + \frac{m^2}{2}\partial_t^{-1} -\frac{1}{8}\, m^{4}\,\partial_{t}^{-3}+ \cdots
\end{equation}
It is ensured by construction that acting twice one some test function with this operator will return the same result as acting once with $\partial_t^2 + m^2$.

In order to see how the definition (\ref{series1}) differs from our Laplace-space definition (\ref{pseudo}) we consider acting with (\ref{series1}) on a function of the form (\ref{lap_x_form}).  The
result is
\begin{equation}
\label{pseudo_series}
  f(\partial_t)\phi(t) = \frac{1}{2\pi i}\oint_C ds\, e^{st}\left[
s + \frac{m^2}{2}s^{-1} -\frac{1}{8}\, m^{4}\,s^{-3}+ \cdots\right] \tilde{\phi}(s)
\end{equation}
The reader will recognize that the series in the square braces
coincides with the Taylor series for $\sqrt{s^2 + m^2}$ about $m=0$.
In the region $|m/s| > 1$ (the high energy part of the phase space)
the series converges to $\sqrt{s^2 + m^2}$.  However, in the region
$|m/s| < 1$ the infinite series fails to converge.  It is due to the
behaviour of the infinite series in this low energy part of phase
space that (\ref{pseudo_series}) and (\ref{pseudo}) do not yield the
same results.  (Notice that the series fails to converge at the
zeroes of the generatrix $s^2=-m^2$ which is not surprising because
these are branch points of the function $\sqrt{s^2 + m^2}$.)

The advantage of the definition (\ref{series1}) over our approach (\ref{pseudo}) is obvious: the series definition preserves the property (\ref{def_id}) which one expects for the square root of an operator.
However, we will argue that our approach seems more sensible from the perspective of constructing quantum field theories.  Since the series (\ref{series1}) fails to converge at $|s| \ll |m|$ this implies
that there is no well-defined low energy limit.  In fact, using this definition, it does not seem possible to recover the usual Klein-Gordon Lagrangian at low energies.  Moreover, the computation of scattering
amplitudes will involve integrating the propagator over all momenta.  Using the definition (\ref{series1}) the low energy part of the phase space integrals will necessarily be ill-defined.  Finally, we point
out a more heuristic problem with the series definition (\ref{series1}).  This definition fails to converge at the curical point $s^2 = -m^2$ and thus it seems necessary to adandon the usual interpretation
of the poles of the propagator as physical states in the associated quantum field theory.

We should stress that our definition (\ref{pseudo}) is just that and
that it may be interesting to consider other definitions such as
(\ref{series1}) which do not lead to the unattractive property which
we have discussed above.    We would also like to stress that that
precisely the same unattractive feature arises for all of the
definitions of fractional differentiation $D^{\alpha}$ which are
studied in the mathematics literature. Thus, it is clear that one
need not view this complication as prohibitive to any potential
physical application of such operators.  The reader who disagrees
with this statement may wish to view the analysis of this subsection
as a motivation to avoid nonlocal theories involving fractional
operators.

\section{Physical Interpretation of the Result}
\label{pde_sec}

\subsection{Linear Partial Differential Equations of Infinite Order}
\label{pde_subsec}

The analysis of section \ref{ode_sec} is actually somewhat more general than what is required in order to study (\ref{padic}, \ref{SFT_intro}).  For differential equations of infinite
order which arise from Lorentz invariant field theories the time derivatives $\partial_t$ must always appear within the d'Alembertian operator
$\Box = -\partial_t^2 + \grad^2$.  Hence we would now like to apply the preceding analysis to linear partial differential equations (PDEs) of infinite order of the form
\begin{equation}
\label{pde}
  F(\Box)\phi(t,{\bf x}) = J(t,{\bf x})
\end{equation}
We refer to the function $F(z)$  in (\ref{pde})
as the \emph{kinetic operator}, which is closely related to the generatrix.
We will be particularly interested in kinetic operators which can be written in the form
\begin{equation}
\label{pde_gen2}
  F(z) = \Gamma(z) \prod_{i=1}^N (-z+m_i^2)
\end{equation}
where $\Gamma(z)^{-1}$ contains no poles in the complex plane, but
is otherwise arbitrary.\footnote{$\Gamma(z)$ should not be confused
with the well-known gamma-function.}  With the ansatz
(\ref{pde_gen2}) equation (\ref{pde}) describes $N$ physical states
with masses $\{m_i\}$. For simplicity we assume the $m_i$ to be
non-degenerate, however, it is simple to drop this restriction.
(Including degenerate masses corresponds to choosing some of the
$r_i$ in equation (\ref{generatrix_ansatz}) to be different from
unity.)

We now proceed to construct the particular solution of (\ref{pde}).
Assuming that the field $\phi(t,{\bf x})$ can be expanded into
Fourier modes as
\begin{equation}
\label{fourier}
 \phi(t,{\bf x}) = \int \frac{d^3k}{(2\pi)^{3/2}} \, e^{i {\bf k} \cdot {\bf x}} \xi_{\bf k}(t)
\end{equation}
equation (\ref{pde}) becomes
\begin{equation}
\label{pde_fourier}
  F(-\partial_t^2 - k^2) \xi_{\bf k}(t) = J_{\bf k}(t)
\end{equation}
where $k^2 \equiv {\bf k} \cdot {\bf k}$, $k \equiv \sqrt{k^2}$ and
\[
  J_{\bf k}(t) = \int \frac{d^3x}{(2\pi)^{3/2}} \, e^{-i {\bf k} \cdot {\bf x} } J(t,{\bf x})
\]
Equation (\ref{pde_fourier}) is of the form (\ref{general_eqn}) where the generatrix is
\begin{eqnarray}
  f(s) &=& F(-s^2 - k^2) \nonumber \\
       &=& \Gamma(-s^2 - k^2) \prod_{i=1}^{N} \, (s + i\omega_k^{(i)} )\, (s - i\omega_k^{(i)}) \label{pde_gen1}
\end{eqnarray}
so that the resolvent generatrix has two poles for each pole of the propagator.  In (\ref{pde_gen1}) we have defined
\begin{equation}
\label{omega}
  \omega_k^{(i)} = \sqrt{ k^2 + m_i^2 }
\end{equation}

Following (\ref{inhom_soln}) we see that the particular solution of (\ref{pde_fourier}) may be written as
\begin{equation}
  \phi_{\bf k}(t) = \frac{1}{2\pi i}\oint_C ds\, e^{st} \frac{\tilde{J}_{\bf k}(s)}{F(-s^2-k^2)}
\end{equation}
so that Laplace-space Green function is $\tilde{G}_k(s) = F(-s^2 - k^2)^{-1}$.
The momentum-space propagator is
\begin{equation}
\label{propagator}
  G(p^2) \equiv \tilde{G}_k(i\omega) = \frac{1}{F(-p^2)}
\end{equation}
with $p^2 \equiv -\omega^2 + k^2$.

We now proceed to solve (\ref{pde}) for the homogeneous case $J=0$.
Since the resolvent generatrix has $2N$ poles (of order one) we expect the solutions to contain $2N$ free coefficients for each $k$-mode (two for each physical
degree of freedom).  It will be convenient to write these $2N$ free coefficients as the real and imaginary parts of $N$ complex numbers $a_k^{(i)}$, $i=1,\cdots,N$.

In solving (\ref{pde}) we wish to apply an additional constraint on the solutions which was not
implied by the analysis of section (\ref{ode_sec}).  Namely, we demand that the solutions $\phi(t,{\bf x})$ be real valued.  The condition $\phi = \phi^{\star}$
translates into the constraint
\begin{equation}
\label{real_fourier}
  \xi_{\bf k}(t)^{\star} = \xi_{-{\bf k}}(t)
\end{equation}
on the Fourier modes.  The general solution of (\ref{pde}) consistent with the reality condition (\ref{real_fourier}) then must take the form
\begin{eqnarray}
  \xi_{ \bf k}(t) &=& \sum_{i=1}^{N} \xi_{\bf k}^{(i)}(t) \label{sum_fourier} \\
  \xi_{\bf k}^{(i)}(t) &=& a_{\bf k}^{(i)} \phi_{\bf k}^{(i)}(t) + a_{\bf -k}^{(i)\star} \phi_{\bf -k}^{\star(i)}(t) \label{real2}
\end{eqnarray}
Each $\xi_{\bf k}^{(i)}$ is the solution corresponding to the $i$-th pole of the propagator (with mass $m_i^2$)
and $\phi_{\bf k}(t)$ are the mode functions (which have been constructed in such a way that $\phi_{\bf k}$ and $\phi_{\bf k}^\star$ are
linearly independent).  The configuration-space solution can be written as
\begin{equation}
\label{config_soln}
  \phi(t,{\bf x}) = \sum_{i=1}^{N}\int \frac{d^3k}{(2\pi)^{3/2}} \left[ a_{\bf k}^{(i)} \phi_{\bf k}^{(i)}(t) e^{i {\bf k} \cdot {\bf x}} + \mathrm{c.c.} \right]
\end{equation}
where ``c.c.'' denotes the complex conjugate of the preceding term.  In the classical solution the coefficients $a_{\bf k}^{(i)}, a_{\bf k}^{(i)\star}$
serve to fix $2N$ initial data $\phi(0,{\bf x}), \dot{\phi}(0,{\bf x}), \cdots, \phi^{(2N-1)}(0, {\bf x})$.  However, in the quantum theory these coefficients are promoted to
annihilation/creation operators.  It is clear that in this context the solution (\ref{config_soln}) describes $N$ physical degrees of freedom,
one for each pole of the propagator.

The solutions $\xi_{\bf k}^{(i)}$ evolve differently depending on the value of $m_i^2$.  There are three qualitatively distinct cases:
\begin{enumerate}

\item \emph{Stable modes:} $m_i^2 > 0$.  This case corresponds to real $\omega_k^{(i)}$.  The solution of (\ref{pde_fourier}) takes the form
(\ref{sum_fourier}, \ref{real2}) with mode function
\begin{equation}
\label{stable_mode_soln}
  \phi_{\bf k}^{(i)} (t) = \frac{e^{-i\omega_k^{(i)}t}}{\sqrt{2 \omega_k^{(i)}}}
\end{equation}
(The normalization of the modes $\phi_k^{(i)}$ is purely conventional since the constant pre-factor may be absorbed into the arbitrary coefficients $a_k^{(i)}$.)

\item \emph{Tachyonic modes:} $m_i^2 \equiv -\mu_i^2 < 0$.  In this case $\omega_k^{(i)}$ is real for $k^2 > \mu_i^2$ and pure imaginary
for $k^2 < \mu_i^2$.  We consider only the latter modes (the instability band) since the former possibility is identical to
the previous case.  For tachyonic modes within the instability band the solution of (\ref{pde_fourier}) takes the form (\ref{sum_fourier}, \ref{real2}) with
mode functions
\begin{equation}
\label{tach_mode_soln}
  \phi_{\bf k}^{(i)}(t) = \frac{1}{2\sqrt{2 \Omega_k^{(i)}}} \left[ e^{\Omega_k^{(i)}t} + i e^{-\Omega_k^{(i)}t} \right]
\end{equation}
where $\Omega_k^{(i)} = \sqrt{\mu_i^2 - k^2}$ is real-valued.

\item \emph{Poles with complex mass}.  Taking $m_i^2$ to be some arbitrary complex numbers
having nonvanishing imaginary part the frequency can be written as
\[
  \omega_k^{(i)} = \alpha_k^{(i)} + i \beta_k^{(i)}
\]
A crucial point is that for a CPT invariant theory the complex-mass states must always arise in conjugate pairs.  Hence
we restrict ourselves to the case where $\omega_k^{(i)\star} = \alpha_k^{(i)} - i\beta_k^{(i)}$ is also a pole.  Real-valued solutions
can be obtained by appropriately superposing the particular solutions $\xi_k$ corresponding to the poles $\omega_k$ and $\omega_k^{\star}$ as
\begin{eqnarray}
  \xi_{\bf k}^{(i)}(t) &=& \frac{e^{\beta_k^{(i)} t}}{\sqrt{2 |\omega_k^{(i)}|}}\left[ a_{\bf k}^{(i)} e^{-i\alpha_k^{(i)} t} + a_{\bf -k}^{(i)\star} e^{+i\alpha_k^{(i)} t}  \right]  \nonumber \\
                        &+& \frac{e^{-\beta_k^{(i)} t}}{\sqrt{2 |\omega_k^{(i)}|}}\left[ b_{\bf k}^{(i)} e^{-i\alpha_k^{(i)} t} + b_{\bf -k}^{(i)\star} e^{+i\alpha_k^{(i)} t}  \right] \label{comp_mode_soln}
\end{eqnarray}
\end{enumerate}

In general the summation (\ref{config_soln}) will contain all three types of modes (\ref{stable_mode_soln}, \ref{tach_mode_soln}, \ref{comp_mode_soln}), although
any particular class could be consistently projected out by altering the prescription for drawing the contour $C$ (see the discussion below equation (\ref{pseudo})).
For example, one could exclude the complex mass solutions by taking $C$ in (\ref{lap_x_form}) to encircle only poles of the resolvent generatrix which lie on either the real axis or the imaginary
axis in the complex plane.  Later on, we will employ this particular choice of $C$ to render both $p$-adic string theory and SFT ghost-free.
Though such a prescription is well defined mathematically, it is not clear how to interpret the resulting truncated theory from a physical perspective.
However, one might imagine taking the contour $C$ as a part of the \emph{definition} of field theory so that different choices of contour yield different theories with different
mass spectra.
Since this prescription restricts the theory to complex frequencies $\omega$ inside $C$ it is, in some sense, analogous to putting a UV cut-off on the
underlying field theory.  This procedure of projecting out unwanted states by suitable choice of $C$ seems reasonable at the level of effective field theory (and hence should
be acceptable for applications to cosmology), however, it is less clear if this is sensible in string theory which is supposed to be UV complete.  On the other hand, notice that the particular
contour which we have discussed still allows for infinitely large frequencies.  Rather, what we are constraining is the \emph{direction} in the complex plane in which the frequency can be made large.

\subsection{Generalization to Curved Backgrounds}
\label{curved_sec}

Though our analysis has been restricted to $3+1$-dimensional flat
Minkowski space, it should be clear that these conclusions readily
generalize to $D$-dimensions.  In curved backgrounds one might
consider generalizing our analysis to equations of the form
\begin{equation}
\label{curved_bkg}
 F\left(\Box_g\right)\phi(t,{\bf x}) = J(t,{\bf x})
\end{equation}
where $\Box_g = g^{\mu\nu}\nabla_{\mu}\nabla_{\nu}$ is the covariant d'Alembertian
operator and we take $F(z)$ of the form (\ref{pde_gen2}).  In the homogeneous case $J=0$
a solution may be constructed by taking
\begin{equation}
\label{curved_soln}
  \phi(t,{\bf x}) = \sum_{i=1}^{N}\phi_i(t,{\bf x})
\end{equation}
where each $\phi_i(t,{\bf x})$ obeys an eigenvalue equation
\begin{equation}
\label{e_value}
  \Box_g \phi_i = m_i^2 \phi_i
\end{equation}
(this approach of taking eigenfunctions of the d'Alembertian was employed to study the cosmology of nonlocal theories in \cite{pi}
and also in \cite{route,lidsey,nongaus}).
Each solution of (\ref{e_value}) (assuming that solutions exist) should admit two initial data
and hence the full solution (\ref{curved_soln}) admits $2N$ initial data, as in our previous flat-space
analysis.  However, it is not so clear if this analysis can be generalized to include inhomogeneous equations
of the form (\ref{curved_bkg}).  Any specific analysis will require detailed knowledge of the pole structure
of the propagator which is, in general backgrounds, a highly nontrivial task.

It is, however, straightforward to generalize our discussion to include homogeneous solutions in de Sitter space.  For a kinetic function
of the form (\ref{pde_gen2}) the de Sitter space generatrix is
\begin{equation}
\label{dS_gen}
  f(s) = \Gamma( -s^2 - 3 H s) \prod_{n} \left(s^2 + 3Hs + m_n^2 \right)
\end{equation}
(the restriction to homogeneous solutions means that we are considering only $k=0$).  There are two zeros for each mass $m_n$
\begin{equation}
\label{dS_zero}
  s_n^{(\pm)} = - \frac{3H}{2}\left[ 1 \pm \sqrt{1-\frac{4 m_n^2}{9 H^2}} \right]
\end{equation}
It is interesting to consider the limit of large Hubble friction $H^2 \gg |m_n^2|$ in which case the solution takes the form
\begin{equation}
\label{dS_soln}
  \phi(t) \cong \phi_0 e^{-3 H t} + \sum_{n} a_n e^{-m_n^2 t / (3 H)}
\end{equation}
In the limit $H \rightarrow \infty$ the first term damps to zero quickly while the second term approaches a constant.  The limit of
large Hubble friction is relevant for studies of inflation in nonlocal theories since during inflation $m^2 / H^2 \sim \mathcal{O}(\eta)$ where $m$
is the characteristic mass scale in the problem and $\eta \ll 1$ is a slow roll parameter.  In this case the almost-constant term in (\ref{dS_soln})
coincides with the slowly rolling solution.

\subsection{Partial Differential Equations Involving Fractional Operators}
\label{frac_pde_subsect}

Using the results of subsection \ref{non_int_root} it is straightforward to consider equations of the form
\[
  (-\Box + m^2)^{n/2}\, \phi(t,{\bf x}) = 0
\]
with $n$ an odd integer.  Assuming that $\phi(t,{\bf x})$ can be
represented as a Fourier integral, we have, in Fourier space,
\begin{equation}
\label{box_n/2}
  \left(\partial_t^2 + k^2 + m^2\right)^{n/2}\, \xi_{\bf k}(t) = 0
\end{equation}
Let us first consider $n=1$ which has only a single mode function.  For stable modes $m^2 > 0$ we have the solution
\begin{equation}
\label{1/2_stable}
  \xi_{\bf k}(t) = (a_{\bf k} + a_{-\bf k}^{\star})\frac{1}{\sqrt{2\omega_k}} J_0(\omega_k t)
\end{equation}
where $\omega_k = \sqrt{k^2 + m^2}$.  (This solution is of the form (\ref{real2}) though it only contains one free coefficient because $\phi_{\bf k}$ and $\phi_{-{\bf k}}$
are not, in this case, linearly independent.)  At late times $\omega_k t \gg 1$ the solution (\ref{1/2_stable}) undergoes damped oscillations
\[
  \xi_{\bf k}(t) \cong (a_{\bf k} + a_{-\bf k}^{\star}) \frac{1}{\omega_k \sqrt{\pi t}} \cos(\omega_k t - \pi / 4)
\]
For tachyonic modes $m^2 = -\mu^2 < 0$ we have the solution
\begin{equation}
\label{1/2_tac}
  \xi_{\bf k}(t) = (a_{\bf k} + a_{-\bf k}^{\star})\frac{1}{\sqrt{2\Omega_k}} I_0(\Omega_k t)
\end{equation}
where $\Omega_k = \sqrt{\mu^2 - k^2}$ and we are assuming that $k < \mu$.  As one would expect the solution (\ref{1/2_tac}) grows exponentially
\[
  \xi_{\bf k}(t) \cong   (a_{\bf k} + a_{-\bf k}^{\star})  \frac{A}{2\Omega_k \sqrt{\pi t}}e^{\Omega_k t}
\]
at late times $\Omega_k t \gg 1$.
The case $m=0$ has been studied in $2+1$-dimensions \cite{canonical_frac} where it has applications
to bosonization \cite{boson}.  (Equations involving fractional powers of the d'Alembertian have also been studied in \cite{frac_dA}.)  This special case admits a canonical quantization
and both causality and Huygens' principle are both respected \cite{canonical_frac}.

We now consider (\ref{box_n/2}) with $n=3$.  In the stable case $m^2 > 0$ the solutions can be cast in the form (\ref{real2}) with mode function
\begin{equation}
  \phi_{\bf k}(t) = \frac{1}{\sqrt{2\omega_k}}\left[ J_0(\omega_k t) + i (\omega_k t) J_1(\omega_k t) \right]
\end{equation}
while, for the tachyonic case $m^2 = -\mu^2 < 0$, we would have
\begin{equation}
  \phi_{\bf k}(t) = \frac{1}{\sqrt{2\Omega_k}}\left[ I_0(\Omega_k t) + i (\Omega_k t) I_1(\Omega_k t) \right]
\end{equation}

It is also straightforward to consider partial differential equations involving fractional derivatives.  Sacrificing Lorentz invariance we might consider the equation
\begin{equation}
\label{frac_pde}
  D^{\alpha} \phi(t,{\bf x}) = \lambda^2 \partial_i \partial^i \phi(t,{\bf x})
\end{equation}
which interpolates between the wave equation (when $\alpha=2$) and the diffusion equation (when $\alpha=1$).  After performing a Fourier transform with respect to the spatial
variables (\ref{frac_pde}) becomes an equation of the form (\ref{frac_ode}).  Thus, for $1 < \alpha \leq 2$ equation (\ref{frac_pde}) admits a well-posed initial value problem
with two initial data and for $ 0 < \alpha \leq 1$ it admits only one piece of initial data.
It is straightforward also to add a mass term to (\ref{frac_pde}).  Equations of this form (and also many more general fractional PDEs, both linear
and nonlinear) have been studied in great detail in the mathematics literature \cite{fractional}.

\subsection{Ghosts and the Ostrogradski Instability}

Though essentially all nonlocal theories containing finitely many higher derivatives have ghosts,
this is not generically true of theories with infinitely many derivatives.
For example, in the case where the propagator contains no poles it is clear from (\ref{config_soln}) that the underlying
field theory has \emph{no} physical excitations at all, ghost or otherwise.  The question of whether or not ghosts are present in an
infinite derivative theory was considered in \cite{tirtho1} where it was shown that the theory will be ghost-free as long
as the propagator contains at most a single pole (of order unity).  However, if the propagator contains two or more poles
then the theory will almost always contain ghost-like excitations.\footnote{Under very special circumstances it is in fact possible to construct ghost-free multi-pole theories \cite{inprog}.  However,
we restrict ourselve to single-pole theories for the present analysis.}  Physically this result is easy to understand: theories with only
a single pole in the propagator describe only one physical degree of freedom and hence the nonlocal structure does not introduce spurious new
ghost states.

If we restrict ourselves to real-valued solutions of linear differential equations which arise from Lorentz-invariant, ghost-free theories with analytic generatrix
then we are limited to equations of the form
\begin{equation}
\label{ghost_free}
  \Gamma(\Box) (-\Box + m^2) \phi(t,{\bf x}) = J(t,{\bf x})
\end{equation}
with $\Gamma(z)$ analytic everywhere in the complex plane.  In the case $J = 0$, as we have previously discussed in subsection \ref{consequence_subsec},
the solutions of this equation are completely insensitive to the choice of $\Gamma(z)$.  In particular, the dynamics for $J=0$ are identical to the solutions of the local wave equation
\[
  (-\Box + m^2)\phi(t,{\bf x}) = 0
\]
so that in the linear, source-free theory the nonlocal structure has
\emph{no effect} on the dynamics (with the possible exception of
re-scaling the quantity which is naively identified as the mass of
the particle).  This result was also observed in \cite{pi}, though
it was not rigorously justified (this result was also observed in
\cite{route}).  Equations of the form (\ref{ghost_free}) fall into
the class of theories referred to as trivially nonlocal in
\cite{woodard2} where it was argued that the nonlocality can be
removed by a field redefinition. (This is, of course, only true at
the linear level.  Inclusion of a nonlinear term in
(\ref{ghost_free}) will spoil the triviality.  It is, unfortunately,
not clear whether such an addition will generically also spoil the
well-posedness of the initial value problem.)

One can also construct ghost-free linear differential equations by considering kinetic functions $F(z)$ with a single zero and also a single pole.  Such equations can be cast
in the form (\ref{ghost_free}) with $\Gamma(z)$ having single pole at some point $z = \mu^2$ and may admit well-posed initial value problem, even at the nonlinear level.  For example,
the equation
\begin{equation}
\label{final_redef}
  f(\partial_t) \phi(t) = U\left[\phi(t)\right]
\end{equation}
with generatrix
\begin{equation}
\label{resum_f}
  f(s) = \frac{2A}{\lambda}\left(1-\frac{s^2}{\lambda^2}\right)^{-1}
\end{equation}
can be seen to be equivalent to the local equation
\begin{equation}
\label{sys1}
    \left(\partial_t^2 - \lambda^2 \right) U\left[\phi(t)\right] + 2 A\lambda \phi(t) = 0
\end{equation}
by acting on both sides of (\ref{final_redef}) with the operator $f(\partial_t)^{-1}$.  One expects the initial value problem for (\ref{sys1})
to be well-posed with only two initial data.
(Similar conclusions apply also to more general equations which involve $f(\partial_t)\phi(t)$ and its first two derivatives.)
The well-posedness of (\ref{final_redef}) is not surprising since
this equation can be obtained from the local theory (\ref{sys1}) by
acting with an inverse differential operator $f(\partial_t)^{-1}$.
This kind of nonlocality is referred to as derived in
\cite{woodard2}.

Finally, we note that it seems to be possible to construct a wide
array of ghost-free theories using fractional operators, as in subsection
\ref{frac_pde_subsect}.  In the case of
equation (\ref{frac_pde}) with $0 < \alpha \leq 2$ we have an example of a nonlocal
theory with well-defined initial value problem which is not
derivable from some local theory, though the price is manifestly broken Lorentz
invariance.  Are equations of the form (\ref{box_n/2})
derivable from local theories?  For an equation of the form
$(-\Box + m^2)^{1-\alpha}\phi = 0$ one might be tempted to define $\phi =
(-\Box + m^2)^{+\alpha} \chi$ so that $\chi$ obeys the local equations
$(-\Box + m^2)\chi = 0$.  However, it is argued in \cite{canonical_frac}
that such a procedure is ill-defined.  Indeed, the transformation
$\phi \rightarrow \chi$ has zeroes in the complex plane and would
not usually be allowed in a field redefinition.  In the case $\alpha = 1/2$
it is clear from (\ref{1/2_stable}) that such a procedure may count incorrectly the
number of initial data.

We have seen how the question of counting initial data is intimately
related to the question of whether the underlying field theory
contains ghosts.  We now consider a related worry: the Ostrogradski
instability \cite{ostrogradski}.  A cartoon of the Ostrogradski
construction follows (for a modern review see
\cite{woodard1,woodard2}). Consider a higher-derivative Lagrangian
which depends nondegenerately on the field and its first $N$
derivatives (so that the equation of motion is of order $2N$ in time
derivatives) the Hamiltonian depends on $2N$ canonical coordinates
corresponding to the $2N$ pieces of initial data which are necessary
to specify the solutions of the Euler-Lagrange equation.
Ostrogradski's theorem states that the Hamiltonian always depends
linearly $N-1$ of the conjugate momenta. It follows that the
Hamiltonian is generically unbounded from below and hence it is
necessarily unstable over half the phase space for large $N$.  In
light of the previous analysis it is easy to see why Ostrogradski's
construction may fail for theories containing \emph{infinitely} many
derivatives.  As long as the propagator only has one pole, then only
two initial data are necessary to specify the solutions of the
equation of motion corresponding to only two independent canonical
coordinates, rather than infinitely many as one would conclude by
naively taking the $N \rightarrow \infty$ limit of Ostrogradski's
result.
For nonlocal theories of infinite order one should instead construct the Hamiltonian using the formalism of \cite{gomis1,gomis2}
which allows for a more transparent identification of the physical phase space of an infinite derivative theory and reduces to the Ostrogradski construction
in the finite derivative case.

It is interesting to note that for theories with more than one pole in the propagator (theories which have ghosts)
turning on large Hubble friction seems to ameliorate the problem somewhat by slowing the evolution of the spurious
modes (see equation (\ref{dS_soln})).  Of course, Hubble friction does not actually \emph{solve} the problem by rendering the Hamiltonian bounded from below,
it merely slows down the higher derivative instability.  Obviously in the case of an infinite spectrum of masses $\{m_n^2\}$ where $|m_n^2| \rightarrow \infty$ as $n \rightarrow \infty$
then one may always find some level $N$ so that $|m_N^2| > H^2$ for finite $H$.

The failure of the Ostrogradski construction for infinite derivative
theories has previously been noted in the math literature, though
not using this language.  In \cite{lalesco} (see also \cite{davis})
it was noted that the naive procedure for writing the $N$-th order
differential equation
\[
  F\left( t, \phi, \frac{d\phi}{dt}, \frac{d^2\phi}{dt^2}, \cdots, \frac{d^{N}\phi}{dt^N}  \right) = 0
\]
as a system of $N$ first order equations
\begin{eqnarray*}
  \frac{d\phi_n}{dt} &=& f_n\left( t, \phi, \phi_1, \phi_2, \cdots, \phi_N \right), \hspace{10mm}n = 1, 2, \cdots, N \\
  \phi_n &=& \frac{d^{n-1}\phi}{dt^{n-1}}
\end{eqnarray*}
fails when $N= \infty$.  Though there exists no general no-go
theorem which excludes the possibility that there exists some
alternate procedure for writing an infinite order ODE as an infinite
system of first order ODEs\footnote{Such a situation would be
reminiscent of multi-Hamiltonian systems where the second order
equation(s) can be written as a first order system in different
ways, depending on the choice of symplectic structure and
Hamiltonian.}, it is very unlikely that such a procedure can be
found.

It is interesting to note that one would obtain quite different conclusions from what we have discussed for the solutions of (\ref{ghost_free})
if one were to truncate the full kinetic function $F(z)$ at some large but finite order in derivatives, $N$.
Consider the truncated kinetic function
\begin{equation}
\label{trunc}
  F_{\mathrm{trunc}}(z) \equiv \sum_{n=0}^N \frac{F^{(n)}(0)}{n!} z^n
\end{equation}
for $N \gg 1$.   Since $F_{\mathrm{trunc}}(z)$ is an $N$-th order
polynomial in $z$ it will inevitably contain some large number of
spurious zeros which were not present in the full kinetic function
$F(z)$. Because the propagator in the truncated theory contains a
large number of spurious poles it follows that the solution
$\phi(t,{\bf x})$ will contain a large number of modes which are not
present in the true theory (\ref{ghost_free}). Inevitably some of
these modes will correspond to ghost-like degrees of freedom so the
solution will be unstable.
Evidently, infinite derivative theories \emph{only} make sense when the full nonlocal
structure of the theory is included.

\section{Some Linear Equations of Physical Interest}
\label{examples_sec}

We now apply the formalism developed in the previous section to some
particular problems of physical interest.  Though we do not uncover
any new solutions, our analysis places previous literature applying
nonlocal field theories to cosmology \cite{pi,lidsey,nongaus} in the
mathematical context developed in this paper, shedding light on the nature
of the initial value problem for these equations.

We proceed by first linearizing the non-linear field equations under
consideration around known solutions, and then solving the
linearized equations using our formal operator calculus.
While we don't derive the error estimates that would make our use of the
linearized equations fully mathematically rigorous, we believe that this
step is justified in the context of the applications of these field
equations to cosmology.  In the case of nonlocal hill-top inflationary models \cite{pi,lidsey,nongaus} the background dynamics which are interesting
for inflation occur very close to the unstable maximum of the potential and the COBE normalization ensures
that inhomogeneities are very small.  Thus it is quite natural to expect that one should obtain accurate results by linearizing the equations of motion about the false
vacuum.  Further support for this approach of perturbing about some constant solution is provided
by the fact that our linearized solution of the $p$-adic string equation (below)
essentially reproduces the leading term (at early times) in the expansion in exponentials used in \cite{zwiebach}.
As we shall see in section \ref{nonlin_sect}, working to higher order in perturbation theory will reproduce the
full solution of \cite{zwiebach}.  Since the solutions of \cite{zwiebach} were verified using a number of non-trivial consistency checks,
we can be fairly confident that errors are small.

\subsection{$p$-adic Strings Near the False Vacuum}
\label{p_false_subsec}

The equation of motion for the tachyon field in $p$-adic string theory is (\ref{padic}).
This equation has the constant solutions $\phi = 0, \pm 1$ for odd $p$, and $\phi = 0, +1$ in the case of even $p$, corresponding to critical points on the
potential.  The solution $\phi = 1$ (and $\phi = -1$ for odd $p$) represents the unstable tachyonic maximum of the potential while $\phi=0$ represents the stable
(true) vacuum of the theory.

We first consider equation (\ref{padic}) linearized about the false
vacuum which physically corresponds to studying small tachyonic
fluctuations which encode the instability of the D$25$-brane in
$p$-adic string theory. Taking
\begin{equation}
\label{p_false}
  \phi(t,{\bf x}) = 1 + \delta \phi(t,{\bf x})
\end{equation}
and linearizing (\ref{padic}) in $\delta \phi$ we find
\begin{equation}
\label{p_eom_false}
  \left[ p^{-\Box / 2} - p \right] \delta \phi(t,{\bf x}) = 0
\end{equation}
The kinetic function
\begin{equation}
\label{p__false_kin}
  F(z) = p^{-z/2} - p
\end{equation}
has infinitely many zeros at $z=z_n$ where
\begin{equation}
\label{zn}
  z_n = -2 + \frac{4\pi i n}{\ln p}
\end{equation}
for $n = 0, \pm 1, \pm 2, \cdots$  (equivalently there are infinitely many poles of the propagator at these points).
Thus the solution of (\ref{p_false}) admits infintely many initial data.  The solution $\delta\phi$ contains a single tachyonic mode with
mass-squared $-2$ in string units.  The behaviour of this mode is described by (\ref{tach_mode_soln}) with $\mu^2=2$.  This is the solution
which would be expected on physical grounds.  However, there are also an infinity of complex-mass solutions of the form (\ref{comp_mode_soln}).
These extra modes have equally spaced mass-squared and, perhaps, correspond physically to the infinite tower of massive states of bosonic string
theory.  It is not clear how reasonable this interpretation is since the (perturbative) Hamiltonian is unbounded from below \cite{gomis2}.\footnote{Recall, however, that
we are perturbing about an unstable maximum of the potential.  Even excluding the complex-mass solutions the \emph{perturbative} Hamiltonian would
appear to be unbounded from below as a result of the tachyonic instability.}  One could render the theory (\ref{padic}) physically sensible by simply projecting
out the unwanted complex-mass poles by taking the contour $C$ in (\ref{lap_x_form}) to enclose only the poles along the real and imaginary axes in complex plane.
We have discussed this prescription previously, at the end of subsection \ref{pde_subsec}.
The fact that (\ref{padic}) can be rendered stable\footnote{Here we mean ``stable'' in the sense of the Ostrogradski instability.  The theory contains a tachyon and hence it is unstable in this sense.}
by suitable choice of $C$ is consistent with \cite{gomis2} where it was shown that the Hamiltonian becomes bounded from below once suitable asymptotic conditions are imposed on the
tachyon field.

\subsection{$p$-adic Strings Near the True Vacuum}

We now consider equation (\ref{padic}) linearized about the true vacuum $\phi = 0$.  Writing
\begin{equation}
\label{p_true}
  \phi(t,{\bf x}) = 0 + \delta \phi(t, {\bf x})
\end{equation}
and linearizing equation (\ref{padic}) in $\delta \phi$ (assuming $p\not= 1$) we have
\begin{equation}
\label{p_eom_true}
  p^{-\Box / 2}\delta \phi(t, {\bf x}) = 0
\end{equation}
The kinetic function
\begin{equation}
\label{p__true_kin}
  F(z) = p^{-z/2}
\end{equation}
has no zeros in the complex plane.  Thus equation (\ref{p_eom_true}) has \emph{no} nontrivial
solution.  This is to be expected on physical grounds since the tachyon vacuum should contain no open string excitations.

This analysis does not rule out the existence of \emph{nonperturbative} solutions in the true vacuum of the theory.  Indeed, \cite{zwiebach}  found anharmonic oscillations about $\phi = 0$
using numerical methods.  The anharmonic oscillations of \cite{zwiebach} appear to contain two free parameters (the frequency $\omega$ and a phase shift which sets the origin of time).

A similar analysis can be performed in the case of open-closed $p$-adic string theory, a simple generalization of (\ref{padic}) which incorporates both the open- and closed-string tachyons.
This analysis is reported on in appendix C.

\subsection{SFT Near the False Vacuum}
\label{SFT_false_sec}

Equation (\ref{SFT_intro}) has two constant solutions: $\phi = -1$ and $\phi=0$.  The former corresponds to the unstable (tachyonic) maximum
while the latter is the true vacuum of the theory.  Writing $\phi = -1 + \delta \phi$ and linearizing in $\delta\phi$ we obtain
\begin{equation}
\label{SFT_true}
  e^{-c \Box}\left[1 + \Box \right] \delta \phi = 0
\end{equation}
so that the kinetic function
\begin{equation}
\label{SFT_kin_true}
  F(z) = e^{-c z} ( 1 + z)
\end{equation}
has only one zero at $z=-1$: exactly the tachyonic mode which is expected on physical
grounds.  It follows that the initial value problem for (\ref{SFT_true}) is well-posed with only two initial data and the solutions are
the tachyonic modes (\ref{tach_mode_soln}) with $\mu^2=1$.

\subsection{SFT Near the True Vacuum}
\label{SFT_true_sec}

We now write $\phi = 0 + \delta \phi$ and linearize (\ref{SFT_intro}) with the result
\begin{equation}
\label{SFT_false}
  \left[ (1+\Box) e^{-c\Box} - 2 \right]\delta \phi = 0
\end{equation}
The kinetic function is
\begin{equation}
\label{SFT_kin_false}
  F(z) = e^{-c z} ( 1 + z) - 2
\end{equation}
The transcendental equation $F(z_n) = 0$ has solutions
\begin{equation}
\label{SFT_zn}
  z_n = -1 - \frac{1}{c} W_n\left( -2 c e^{-c} \right)
\end{equation}
where $W_n$ are the branches of the Lambert W-function.  It is straightforward to check that for $c = \ln (3^3/4^2)$ none of the $z_n$ are real so that (\ref{SFT_false}) admits infinitely
many initial data.  The solutions are the complex-mass states (\ref{comp_mode_soln}) which, perhaps, can be physically interpreted as closed string excitations (at large $n$ equation (\ref{SFT_zn}) describes
a spectrum with equally spaced masses-squared).  However, such an interpretation should be employed with great care since the Hamiltonian for this system is unbounded from below \cite{gomis2}.  See also
\cite{null} for further discussion.

Notice that, as in subsection \ref{p_false_subsec}, we could project out the unwanted complex-mass states by taking the contour $C$ in (\ref{lap_x_form}) to enclose only the real and imaginary
axes in the complex plane.

\section{Nonlinear Equations of Infinite Order}
\label{nonlin_sect}

\subsection{The Nonlinear Problem}
\label{discussion_sec}

A systematic treatment of nonlinear differential equations of infinite order is beyond the scope of this paper (indeed, such a treatment is absent even for the case of finite order).
Here we discuss some promising approaches and, in the next subsection, we will consider a perturbation theory approach to the nonlinear equation of $p$-adic
string theory, rendering the initial value problem physically sensible by suitable choice of the contour of integration, $C$.  Finally, in subsection \ref{redef_implications} we will
discuss the implications of this choice of $C$ on the full nonlinear problem.

A powerful and efficient method of generating solutions for infinite order nonlinear equations (which can be used also in cosmology) is the iterative procedure
\cite{nl_cosmo,Joukovskaya,iterative1,iterative2}.  Iterative techniques are quite common in the mathematical literature on nonlinear integral equations and the
appeal of such procedures for equations of the form (\ref{gen_intro}) is not surprising given the relationship detailed in appendix A.  For the purposes of characterizing
the initial value problem, however, such techniques do not seem to be the most appropriate since they typically rely on some pre-existing understanding of the asymptotic behaviour of
the solutions.  (One advantage of the perturbative approach described in the next subsection is that one can specify the field and its derivatives at $t=0$ and allow the system to evolve to
a unique solution at $t=+\infty$.)  On the other hand, one might nevertheless argue that iterative approaches do still shed light on the initial value problem.  This is so because
for certain stringy models there is numerical evidence that the iterative procedure always converges to a unique solution once the sign at infinity and the value of the field at $t=0$ have been
specified \cite{iterative2}.  This suggests a class of theories of the form (\ref{gen_intro}) with only two arbitrary numbers in the solution.

Another promising approach is the connection to the heat equation \cite{singular,route,vlad} which may also be used to shed light on the nature of the initial value problem
\cite{route}.  However, it is not clear if this approach can also be employed for more general kinetic functions which do not involve $e^{-\alpha\Box}$.  In \cite{route} solutions in
some stringy models were constructed which depend only on a finite number of initial data.

Finally, we should also mention the expansion in eigenfunctions of the kernel which was employed in \cite{vlad}
for the case of $p$-adic string theory and which does not seem to have recieved as much attention as the alternative approaches.  The expansion in Hermite polynomials of \cite{vlad} seems
to provide a very promising approach.  Like the perturbative approach which we will describe in the next subsection, this expansion is systematic and can, in principle, be continued to arbitrarily
high accuracy.  However, unlike the perturbative approach, it does not rely on any a priori assumptions about the initial data.

At this point we would like to comment upon an apparent discrepency in the literature.  How does one reconcile the results of \cite{woodard2}
(also subsections \ref{p_false_subsec} and \ref{SFT_true_sec}) that equations (\ref{padic}, \ref{SFT_intro}) admit infinitely many initial data with evidence (coming from the techniques described
above) that these equations admit only finitely many initial data?  These results are not necessarily in conflict because many previous studies of equations of the form
(\ref{gen_intro}) may have (either implicitly or explicitly) placed constraints on the  solutions which omit the ghost-like states.  For example, it was shown in \cite{gomis2} that the Hamiltonian for both
$p$-adic string theory and also SFT becomes bounded from below if one demands that the tachyon is at the false vacuum as $t\rightarrow -\infty$.  Thus, the ghost-like excitations
will not be present in any studies which make this assumption.\footnote{The fact that both $p$-adic string theory
and SFT are rendered ghost-free once this auxiliary constraint is imposed means that the ghost-like modes are not present when one considers the specific physical problem of studying brane decay
in these theories.  However, this resolution seems somewhat unsatisfying since one would like to be able to consider such nonlocal theories in a more general setting.  The prescription for re-defining
the pseudo-differential operator discussed in subsection \ref{pde_subsec} provides an auxiliary constraint which also projects out the ghost-like states, but is less restrictive than demanding
that $\phi$ sit at the unstable maximum as $t\rightarrow -\infty$.}
Finally, it was noted in \cite{woodard2} that perturbation theory projects our the ghost-like modes in the case of string field theory (this is so because in the case of equation (\ref{SFT_intro}) the nonlocal structure
of the theory can be put into the interaction term by a field re-definition) which explains why path integral quantizations \cite{quantize} do not exhibit any signs of pathology.  (See also \cite{modular} for details
of the Feynman rules in string field theory.)

\subsection{Perturbative Approach}
\label{nl_sub1}

Consider the full nonlinear equation of $p$-adic string theory (\ref{padic}) with initial conditions close to $\phi = 1$.
For simplicity we consider only the homogeneous case $\phi = \phi(t)$ (inhomogenous solutions in $p$-adic string theory
have been considered in \cite{mode,caustics}).  In this case it is appropriate to expand the $p$-adic scalar in perturbation theory as
\begin{equation}
\label{pert_expand}
  \phi(t) = 1 + \sum_{n=1}^{\infty}\frac{1}{n!}\delta^{(n)}\phi(t)
\end{equation}
It is straightforward perturb the field equation (\ref{padic}) up to second order with the result
\begin{eqnarray}
  \left[ p^{\partial_t^2/2} - p \right] \done \phi &=& 0 \label{lin_eom}\\
  \left[ p^{\partial_t^2/2} - p \right] \dtwo \phi &=& p(p-1)\left(\done \phi \right)^2
  \label{sec_eom}
\end{eqnarray}

In general, for the $n$-th order perturbation, one will obtain an equation of the form
\begin{equation}
\label{n_pert}
  \left[ p^{\partial_t^2/2} - p \right] \delta^{(n)} \phi(t) = J_n(t)
\end{equation}
where $J_n(t)$ is constructed from perturbations of order less that $n$.
At each order in perturbation theory the generatrix has the form
\begin{equation}
\label{p_pert_gen}
  f(s) = p^{s^2/2} - p
\end{equation}
We project out the complex-mass states (discussed in subsection \ref{p_false_subsec}) by taking the contour $C$ to encircle only poles on the real and imaginary axes in the
complex plane.  Thus we are free to write (\ref{p_pert_gen}) in the form
\begin{equation}
\label{p_gen_ansatz}
  f(s) = \gamma(s) (s + \sqrt{2} ) (s - \sqrt{2} )
\end{equation}
with $\gamma(s) = \left[ p^{s^2/2} - p \right] / \left[s^2 - 2\right]$ having no zeros in the domain of interest.
Therefore, at each order in perturbation theory there are only two initial conditions and hence that the full nonperturbative solution
also admits only two initial conditions.  For the sake of illustration we choose $\phi(0) = 1 + \epsilon$ and $\dot{\phi}(0) = 0$
where we assume that $|\epsilon| \ll 1$.

Inspection of (\ref{n_pert}) shows that the full solution will have the form of a sum of exponentials
\[
  \phi(t) = \sum_{n=-\infty}^{+\infty} a_n e^{\sqrt{2} n t}
\]
similar to what was employed in \cite{zwiebach} and \cite{pi,nongaus}.  (In previous applications the terms with $n < 0$ were omitted by the boundary condition
$\phi(-\infty) = 1$, here we keep those terms since we wish to impose our initial conditions at $t=0$.)

We now proceed to solve the perturbation equations.  The linear perturbation (\ref{lin_eom}) is straightforward to solve using (\ref{hom_soln})
\begin{equation}
\label{lin_pert_soln}
  \done \phi(t) = \epsilon \cosh \left( \sqrt{2} t \right)
\end{equation}
which obeys $\done\phi(0) = \epsilon$, $\done\dot{\phi}(0) = 0$.
The second order equation (\ref{sec_eom}) takes the form
\begin{eqnarray}
  \left[ p^{\partial_t^2/2} - p \right] \dtwo \phi &=& J_2(t) \\
  J_2(t) &=& \frac{\epsilon^2}{2}p (p-1) \left[ \cosh(2\sqrt{2}t) + 1 \right]
\end{eqnarray}
In Laplace space the source $\tilde{J}_2(s)$ can be written as
\begin{equation}
  \tilde{J}_2(s) = \frac{\epsilon^2}{4}p (p-1) \left[ \frac{1}{s-2\sqrt{2}} + \frac{1}{s+2\sqrt{s}} + \frac{2}{s} \right]
\end{equation}
Plugging this into (\ref{inhom_soln}) and performing the contour integration yields the particular solution
\begin{equation}
\label{pert_part}
  \dtwo\phi_p(t) = \frac{\epsilon^2(p-1)}{2(p^3-1)}\cosh(2\sqrt{2}t) + \frac{\epsilon^2(p-1)}{3\ln p}\cosh(\sqrt{2}t) - \frac{\epsilon^2 p }{2}
\end{equation}
To (\ref{pert_part}) we are free to add a solution of the homogeneous equation in order to fix the initial conditions.  The appropriate choice is
\begin{equation}
  \dtwo\phi_h(t) = -\epsilon^2 \left[ \frac{p-1}{2(p^3-1)} + \frac{p-1}{3\ln p} - \frac{p}{2} \right]\cosh(\sqrt{2}t)
\end{equation}
The full second order solution $\dtwo\phi(t) = \dtwo\phi_p(t) + \dtwo\phi_h(t)$  obeys $\dtwo\phi(0) = \dtwo\dot{\phi}(0) = 0$ and can be written as
\begin{equation}
  \dtwo\phi(t) = \frac{\epsilon^2(p-1)}{2(p^3-1)} \left[  \cosh(2\sqrt{2}t) - \cosh(\sqrt{2}t) \right] + \frac{\epsilon^2 p}{2}\left[ \cosh(\sqrt{2}t) - 1 \right]
\end{equation}
In principle this procedure could be continued up to arbitrarily high order in perturbation theory.
It should be clear the the perturbative method employed in this section could also be readily applied to the SFT equation of motion (\ref{SFT_intro}) or to other nonlinear equations of the
form (\ref{gen_intro}).




\subsection{The Implications of Restricting the Contour of Integration}
\label{redef_implications}

At the end of subsection \ref{pde_subsec}  we have advocated
re-defining the action of the formal pseudo-differential operator
$f(\partial_t)\phi(t)$ in (\ref{pseudo}) taking $C$ to only enclose
the real and imaginary axes in the complex plane.  We have shown in
section \ref{examples_sec} that this prescription renders both the
$p$-adic string and also SFT ghost-free (at the linearized level) by
projecting out the unwanted complex-mass states.  In subsection
\ref{nl_sub1} we have illustrated how this approach may be employed
order-by-order in perturbation theory so that one could (in
principle) construct solutions of the full nonlinear problem
(\ref{padic}) which depend on only two initial data (it should be
clear that the same approach will also work for SFT
(\ref{SFT_intro})).  However, there is nothing inherently
perturbative about the definition (\ref{pseudo}) and one could
imagine imposing this restriction even beyond perturbation theory.
Our prescription therefore provides a nonperturbative means of
re-defining both $p$-adic string theory and SFT in such a way as to
evade the Ostrogradski instability.\footnote{The idea of finding
some nonperturbative means of restricting SFT to a naive subclass of
its solutions in order to evade the Ostrogradski instability was
suggested in \cite{woodard2}.}

However, one may wonder what the implications of this restriction
are for the fully nonlinear problem. It would be a shame if this
re-definition were to project out not only the ghosts but also some
physically interesting nonperturbative solutions. We now investigate
this question, showing that our prescription does not affect most of
the interesting solutions of (\ref{padic}, \ref{SFT_intro}) which
have been considered in the literature. We will focus primarily on
$p$-adic string theory, though it will be clear that similar
conclusions will also apply for SFT.

First, we note that at the nonlinear level, one could think about this restriction on $C$ in (\ref{pseudo})
as limiting ourselves only to look for solutions $\phi(t)$ within a certain class of functions.  Specifically, if
$\phi(t)$ can be written as (\ref{lap_x_form}) with $C$ only enclosing the real and imaginary axes, then clearly the action
of $f(\partial_t)$ is unaltered by our prescription.  (This is simply to say that if none of the poles $\{s_i\}$ of $\tilde{\phi}(s)$
have both $\mathrm{Re}(s_i)$ and $\mathrm{Im}(s_i)$ non-zero then an infinite contour $C$ in (\ref{pseudo}) can be deformed to only encircle the real and imagninary axes.)
This restriction clearly defines a rather large class of functions - it admits any function which may be represented as a sum (either discrete or continuous) of exponential
modes of the form $e^{i \omega t}$, $e^{\lambda t}$ with $\omega$, $\lambda$ real.  In particular, this restriction does not omit any function which has a well-defined Fourier
transform (since such functions can be represented as a sum of modes $e^{i \omega t}$).  Indeed, this restriction is considerably \emph{less stringent} than demanding that the
solution $\phi(t)$ has a Fourier transform since it also admits a wider class of function (such as $e^t$) which do not.

Let us consider some particular nonperturbative solutions of (\ref{padic}) which have been constructed in the literature.  The anharmonic oscillations of \cite{zwiebach} (which have been suggested to represent
closed string excitations) can be written as a discrete Fourier series and hence this class of solutions is not omitted.
Equation (\ref{padic}) admits a kink-like solution \cite{padic_st,zwiebach,padic_math,vlad} which interpolates between the unstable maxima $\phi = \pm 1$ at $t \rightarrow \pm \infty$ (for odd $p$).  Since this
kink solution can be thought of as the zero-frequency limit of the anharmonic oscillations \cite{zwiebach} it follows that this solution is also untouched by our restriction. Similar comments apply for kink-type solutions
in SFT and other stringy models \cite{phantom1}-\cite{phantom4}, \cite{nl_cosmo,Joukovskaya,iterative1,iterative2}.  The rolling solution of \cite{zwiebach} (and all solutions which can be
constructed as summations of the form $\sum_n e^{n\lambda t}$, both in $p$-adic string theory and in SFT \cite{schnabl}-\cite{comments}) also remains unexcluded.
Finally, note that equation (\ref{padic}) admits the rapidly growing solution
\begin{equation}
\label{grow}
  \phi(t) = p^{\frac{1}{2(p-1)}}\exp\left(\frac{1}{2}\frac{p-1}{p\ln p}t^2\right)
\end{equation}
(see, for example, \cite{vlad,caustics}).  The identity
\[
  \int_{-\infty}^{+\infty} d\omega \, e^{-a\omega^2 - \omega t} = \sqrt{\frac{\pi}{a}} \, e^{t^2 / (4a^2)}
\]
implies that this solution can be represented as a sum of
exponential functions with real-valued argument and hence this
solution also survives our restriction.  (Rotating $t \rightarrow i
t$ we see that any gaussian lump-type solutions are also
unexcluded.) One the other hand, clearly our prescription will
exclude solutions which involve summations of terms of the form
(\ref{comp_mode_soln}), such as the infinite-parameter solution of
(\ref{padic}) described by equation (88) of \cite{gomis2}.  We see, then,
that the restriction which one must impose to render the theories (\ref{padic}, \ref{SFT_intro})
ghost-free is not terribly onerous.

An interesting and important question is whether this prescription preserves the perturbative S-matrix.  Though we do not have any rigorous proof, we expect that it does as long
as one only considers S-matrix elements involving states with $m^2$ real-valued (which, in any case, are the physically meaningful states).  This agrument is supported by evidence from
perturbative string field theory.  As we have seen in subsection \ref{SFT_false_sec}, our prescription has no effect on SFT if one considers a perturbative expansion about the unstable maximum.
(This is so because perturbation theory also projects out the unwanted complex-mass states \cite{woodard2}.)  Since perturbative string field theory is believed to be consistent \cite{quantize}
and to reproduce the usual string theory amplitudes \cite{modular} it seems clear that, at least in this particular case, our prescription does not unacceptably modify the S-matrix.  Of course,
this argument is far from conclusive, and it is an interesting problem study the implications of our prescription on the S-matrix in a more general setting.  We leave the complete resolution of
this issue to future investigations.


\section{Conclusions}
\label{concl}

In this paper we have presented a simple and intuitive formalism for
studying the initial value problem
in a wide variety of nonlocal theories.  Contrary to
naive arguments, differential equations of infinite order do not
necessarily admit infinitely many initial data.  Rather, we have
shown that every pole of the propagator contributes two initial data
to the final solution.  Crucially, we have shown that this counting procedure exhausts \emph{all}
possible solutions at the linear level.  This result has a transparent physical
interpretation since each pole of the propagator should correspond
to a physical excitation and the two initial data per physical state
are promoted to annihilation/creation operators in the quantum
theory.

We have considered, in particular, the dynamical equations of $p$-adic string
theory and string field theory arguing that in both cases the initial value problem
admits infinitely many initial data.  However, both theories may be rendered ghost-free
by suitably re-defining the action of the pseudo-differential operator
(taking the contour in (\ref{pseudo}) to enclose only the real and imaginary axes)
and we have suggested that such a re-definition may be analogous to putting a UV cut-off on the theory.
However, this procedure lacks firm mathematical motivation and may be
difficult to interpret physically.  Indeed, a skeptic might argue that it is
tantamount to simply excluding the unwanted solutions by hand.

On the other hand, we have argued that one might take the contour $C$ as part of the definition of the nonlocal theory so
that different choices of $C$ yield different theories with different mass spectra.  In \cite{woodard2} it was suggested that one
might evade the Ostrogradski instability in string field theory by somehow re-defining the theory to constrain it to a naive subclass of
its solutions.  It was noted there that defining the theory through a perturbative expansion (analogous to the analysis in subsection \ref{SFT_false_sec}) accomplishes just this.
Our prescription for redefining the action of the pseudo-differential operator accomplishes the same thing, though it is non-perturative (there is nothing inherently perturbative
about (\ref{pseudo}) and, indeed, an analogous definition is used to study nonlinear equations in \cite{analysis}).  The same prescription works for both $p$-adic string theory
and SFT.  Moreover, in both cases this prescription seems to preserve the nice features of the theory.  We have shown in
subsection \ref{redef_implications} that this restriction does not exclude most of the interesting solutions which have been constructed previously in the literature.
We believe that this is an idea which merits further investigation.

One limitation of our analysis is that we have, for the most part, restricted our attention to linear differential equations.  Though this certainly omits some interesting problems, it is sufficient
for many physical applications since a great deal of information can be extracted by perturbing about some vacuum solution.  We have illustrated this by studying the dynamical equation of $p$-adic
string theory up to second order in perturbation theory in subsection \ref{nl_sub1}.  Since, at least in principle, one could construct nonperturbative solutions by resumming the perturbation series we expect that a similar
counting of initial data should apply also for nonlinear equations.  However, we leave a detailed study of the solutions of nonlinear equations of infinite order to future investigations.

It is sometimes argued that obtaining any continuum field theory of quantum gravity will require abandoning locality \cite{woodard2}.  Since
presumably any viable nonlocal theory should admit only finitely many initial data (at most two if we wish to evade the Ostrogradski instability) it is therefore an interesting question
as to whether one can construct interacting fundamentally nonlocal theories which admit only finitely many initial data
(preferably without having to resort to re-defining the action of the pseudo-differential operator).
This is certainly possible, the model of \cite{maximal} provides an explicit example (see also discussions in \cite{IVP,woodard1,woodard2}).
We have noted that equations involving fractional operators also seem to provide a promising class of theories.  We believe that the formalism which we have presented
in this paper will be useful in finding also more general classes of theories which satisfy these criterion.

\section*{Note Added}

After completing this paper we became aware of work by Lee and Wick \cite{LW} and also by Cutkosky, Landshoff, Olive and Polkinghorne \cite{clop} which may have some bearing on the our prescription
for re-defining the formal pseudo-differential operator.  In \cite{LW} Lee and Wick proposed a modification of electrodynamics which involves finitely many higher derivatives.  The Lee-Wick model
has recently been revived as a possible solution to the hierarchy problem \cite{LWSM}.   In this construction the propagator for each standard model (SM) field contains two poles: one corresponding
to the usual SM state and the other corresponding to an associated Lee-Wick particle.  In \cite{clop} a modification of the usual contour prescription for Feynman diagrams was
proposed which preserves the unitarity of the
theory and removes the exponential growth of disturbances associated with the Ostrograski instability coming from the Lee-Wick parteners.  The modification of
\cite{clop} seems similar in spirit to our suggestion for deforming the contour of integration in the definition of the action of the formal pseudo-differential on Laplace space.  The prescription employed in
\cite{clop} can be shown to be equivalent to imposing a future boundary condition that forbids outgoing exponentially growing modes and leads to violations of causality.  These violations of causality are restricted to
microscopic scales \cite{coleman} and the
theory is believed to respect macroscopic causality and to be free of paradoxes.  Though it remains to be seen whether similar statements can be made about the prescription which we have advocated, the
consistency of the Lee-Wick model seems to provide a good reason to be optimistic on this front.  It would be interesting to investigate this issue in detail.

\section*{Acknowledgments}

This work was supported in part by NSERC. We are grateful to T.\
Biswas for interesting and enlightening discussions and also to J.\
Cline, K.\ Dasgupta, and S.\ Prokushkin for valuable comments on the
manuscript. It is also a pleasure to thank I.\ Ellwood, N.\ Jokela,
M.\ Jarvinen, E.\ Keski-Vakkuri and J.\ Majumder for interesting
correspondence.  Finally, we are indebted to G.\ Calcagni, J.\
Gomis, L.\ Joukovskaya, R.\ Woodard and B.\ Zwiebach for comments.

\renewcommand{\theequation}{A-\arabic{equation}}
\setcounter{equation}{0}  

\section*{APPENDIX A: Convolution Form of the $p$-adic Equation}

The relationship between equation (\ref{padic}) and a certain nonlinear convolution equation
has previous been noted in the literature \cite{zwiebach}.  Using (\ref{pseudo}) the quantity $p^{\partial_t^2 / 2}\phi(t)$
on the right-hand-side of (\ref{padic}) can be written as
\begin{eqnarray*}
  p^{\partial_t^2/2}\phi(t) &=& \frac{1}{2\pi i}\oint_C ds\, p^{s^2/2} \tilde{\phi}(s) e^{st} \\
  &=& \frac{1}{2\pi i}\oint_C ds\, \tilde{\phi}(s) e^{st} \left[\frac{1}{\sqrt{2\pi \ln p}}\int_{-\infty}^{+\infty}dt' \, e^{-(t-t')^2/(2\ln p)}e^{-s(t-t')}\right]  \\
  &=& \frac{1}{\sqrt{2\pi \ln p}}\int_{-\infty}^{+\infty}dt'\, e^{-(t-t')^2/(2\ln p)}\left[ \oint_C ds\, \tilde{\phi}(s) e^{st'} \right] \\
  &=&  \frac{1}{\sqrt{2\pi \ln p}} \int_{-\infty}^{+\infty} dt'\, e^{-(t-t')^2 / (2 \ln p)} \phi(t')
\end{eqnarray*}
Then equation (\ref{padic}) is equivalent to the following nonlinear Fredholm equation
\begin{equation}
\label{convolution}
  \frac{1}{\sqrt{2\pi \ln p}} \int_{-\infty}^{+\infty} e^{-(t-t')^2 / (2 \ln p)} \phi(t') = \phi(t)^p
\end{equation}
which gives the convolution form of (\ref{padic}).  (See, for example, \cite{zwiebach,padic_math,vlad} for details.)

The above derivation relied heavily on the gaussian form of the kinetic function in equation (\ref{padic}).
The action of kinetic operators involving more general functions may also be written as a convolution with some kernel.
To see this, we write
\begin{equation}
\label{int_eqn}
  \int_{-\infty}^{+\infty}dt' \,  K(t')\phi(t+t') = f(\partial_t) \phi(t)
\end{equation}
If we assume that the integrand of (\ref{int_eqn}) may be expanded as
\begin{equation}
\label{app2}
  \phi(t+t') = \sum_{n=0}^{\infty}\frac{(t')^n}{n!}\phi^{(n)}(t)
\end{equation}
then, employing the series expansion for the generatrix (\ref{generatrix_series}), we see that both sides of (\ref{int_eqn}) match for
\begin{equation}
\label{app3}
  f^{(n)}(0) = \int_{-\infty}^{+\infty}dt \, t^n  K(t)
\end{equation}

\renewcommand{\theequation}{B-\arabic{equation}}
\setcounter{equation}{0}  

\section*{APPENDIX B: Review of Laplace Transforms}

We define a function $f(t)$ and its Laplace transform $\tilde{f}(s)$ by the transformations
\begin{eqnarray}
  \phi(t) &=& \mathcal{L}^{-1} \{\tilde{\phi}(s)\} =  \frac{1}{2\pi i} \int_{a - i \infty}^{a + i\infty} e^{st} \tilde{\phi}(s)ds \label{forward} \\
  \tilde{\phi}(s) &=& \mathcal{L}\{\phi(t)\} = \int_{0}^{\infty} e^{-st} \phi(t)dt \label{backwards}
\end{eqnarray}
The $ds$ integration in (\ref{forward}) is performed along a vertical line in the complex $s$-plane and $a$ should be
chosen sufficiently large that all poles of the integrand are to the left of the contour.  For $t > 0$ we
can close the contour to the left using an infinite semi-circle so that
\[
  \phi(t) = \frac{1}{2\pi i} \oint_{C} e^{st} \tilde{\phi}(s)ds \hspace{5mm}\mathrm{for}\hspace{5mm}t>0
\]
which gives equation (\ref{lap_x_form}).  On the other hand, for $t < 0$ the contour should be closed to the right and the integration
gives zero because the integrand is everywhere analytic within the contour of integration.  Strictly speaking (\ref{backwards}) applies
only for $\mathrm{Re}(s) > a$.  For $\mathrm{Re}(s) < a$ the Laplace-space function $\tilde{\phi}(s)$ should be defined by analytic continuation.

Differentiation acts in Laplace space as
\begin{equation}
\label{app_der}
  \partial_t^{(n)}\phi(t) = \frac{1}{2\pi i}\oint_C ds\, e^{st} \left[ s^n \tilde{\phi}(s) - \sum_{i=1}^n d_j s^{n-j}  \right]
\end{equation}
where
\begin{equation}
\label{dj_app}
  d_j = \partial_t^{(j-1)}\phi(0)
\end{equation}
and $\sum_{j=1}^0$ is identically zero.  Following \cite{analysis} one might define the action of $f(\partial_t)\phi(t)$, with $f(s)$ given by (\ref{generatrix_series}), as
\begin{equation}
\label{pseudo_app}
  f(\partial_t)\phi(t) = \frac{1}{2\pi i}\oint_C ds\, e^{st} \left[ f(s) \tilde{\phi}(s) - \sum_{n=0}^{\infty} \sum_{i=1}^n \frac{f^{(n)}(0)}{n!} d_j s^{n-j}  \right]
\end{equation}
As discussed in section \ref{ode_sec}, in solving (\ref{general_eqn}) the second term in (\ref{pseudo_app}) may be adsorbed into the arbitrary coefficients in the solution $\tilde{\phi}(s)$.
This is material is discussed, for example, in \cite{analysis,lap_refs}.

\renewcommand{\theequation}{C-\arabic{equation}}
\setcounter{equation}{0}  

\section*{APPENDIX C: Open-Closed $p$-adic Strings}

A simple generalization of open $p$-adic string theory (\ref{padic}) which incorporates also the closed string tachyon is the
coupled infinite order system \cite{open-closed,mode}
\begin{eqnarray}
  p^{-\Box / 2}\phi &=& \phi^p \psi^{p(p-1)/2} \label{open} \\
  p^{-\Box / 4}\psi &=& \psi^{p^2} + \frac{\lambda^2(p-1)}{2p}\psi^{p(p-1)/2 - 1}\left(\phi^{p+1}-1\right) \label{closed}
\end{eqnarray}
where $\phi$ represents the open-string tachyon, $\psi$ represents the closed string tachyon and $\lambda^2 \ll 1$ is related to the string coupling.
Let us put the open-string tachyon in its vacuum state $\phi = 0$ and consider the resulting closed-string tachyon dynamics.  For $\phi=0$ equation (\ref{closed}) becomes
\begin{equation}
\label{closed2}
  p^{-\Box / 2}\psi = \psi^{p^2} - \frac{\lambda^2(p-1)}{2p}\psi^{p(p-1)/2 - 1}
\end{equation}

This equation admits two constant solutions $\psi = \psi_f$ and $\psi = \psi_t$ representing the false and true vacuum states respectively.  The false vacuum is
\begin{equation}
\label{psif}
  \psi_f = 1 + \frac{\lambda^2}{2p(p+1)} + \mathcal{O}(\lambda^4)
\end{equation}
Writing $\psi = \psi_f + \delta\psi$ and linearizing (\ref{closed2}) in $\delta\psi$ gives
\begin{equation}
\label{open-closed-false}
  \left[ p^{-\Box / 4} - \left( p^2 + \lambda^2\frac{p^3 + p -2}{4p}\right) \right]\delta \psi = 0
\end{equation}
to leading order in $\lambda^2$.  The kinetic function
\begin{equation}
\label{kin_closed_false}
  F(z) = p^{-z / 4} - \left( p^2 + \lambda^2\frac{p^3 + p -2}{4p}\right)
\end{equation}
has infinitely many zeros at $z = z_n$ where
\begin{equation}
  z_n = -\frac{4}{\ln p}\ln\left[p^2 + \frac{\lambda^2(p^3+p-2)}{4p}\right] + \frac{8\pi i n}{\ln p}
\end{equation}
and $n = 0, \pm 1, \cdots$

We consider now fluctuations about the true vacuum $\psi_t$.  For $p > 2$ (the case $p=2$ is treated below) we have
\begin{equation}
\label{psit1}
  \psi_v = 0
\end{equation}
and writing $\psi = 0 + \delta\psi$ yields
\begin{equation}
\label{open-closed-true1}
  p^{-\Box / 4} \delta \psi = 0
\end{equation}
The kinetic function $F(z) = p^{-z/4}$ has no zeros and hence there is no solution.  For $p=2$ the true vacuum is shifted by the interaction
\begin{equation}
\label{psit2}
  \psi_v = -\frac{\lambda^2}{4} + \mathrm{O}(\lambda^6)
\end{equation}
however, writing $\psi = -\lambda^2 /4 + \delta\psi$ yields again $F(z) = p^{-z/4}$.

One might instead imagine writing $\phi = 1 + \delta\phi$ and $\psi = 1 + \delta \psi$ in which case the system of equations (\ref{closed},\ref{open})
gives
\begin{eqnarray}
  \left[ p^{-\Box / 2} - p \right]\delta\phi &=& \frac{p}{2}(p-1)\delta \psi \label{open_dpert} \\
  \left[ p^{-\Box / 4} - p^2 \right]\delta \psi &=& \frac{\lambda^2 (p^2-1)}{4p}\delta\phi  \label{closed_dpert}
\end{eqnarray}
For $\lambda^2 = 0$ the study of this system is very similar to the analysis in section \ref{nonlin_sect}.  For finite $\lambda^2$
solutions are also possible and have been described in \cite{mode}.

\end{document}